\documentclass{aastex6}

\shorttitle{Formation of plasmoid chains during evolution of the tilt instability}
\shortauthors{Baty et al.}

\usepackage{bm}%
\usepackage{graphicx}%
\usepackage{amsmath}

\def\ltsima{$\; \buildrel < \over \sim \;$}
\def\gtsima{$\; \buildrel > \over \sim \;$}
\def\simlt{\lower.5ex\hbox{\ltsima}}
\def\simgt{\lower.5ex\hbox{\gtsima}}

\begin{document}

\title{Formation of plasmoid chains and fast magnetic reconnection during nonlinear evolution of the tilt instability}

\email{hubert.baty@unistra.fr}

\author{ Hubert BATY}
\affiliation{Observatoire Astronomique de Strasbourg, Universit\'e de Strasbourg, CNRS, UMR 7550, 11 rue de l'Universit\'e, F-67000 Strasbourg, France}

\begin{abstract}

We investigate, by means of two-dimensional incompressible magnetohydrodynamic (MHD) numerical simulations,
the fast collisional magnetic reconnection regime that is supported by the formation of plasmoid chains when the Lundquist
number $S$ exceeds a critical value (at magnetic Prandtl number, $P_m =1$). A recently developed characteristic-Galerkin finite-element code,
FINMHD, that is specifically designed for this aim in a reduced visco-resistive MHD framework, is employed. 
Contrary to previous studies, a different initial setup of two repelling current channels is chosen in order to form two
quasi-singular current layers on an Alfv\'enic time scale as a consequence of the tilt instability.
If $S \simlt 5 \times 10^3$, a subsequent stationary reconnection process
is obtained with a rate scaling as $S^{-1/2}$ as predicted by the classical Sweet-Parker model.
Otherwise, a stochastic time-dependent reconnection regime occurs, with a fast time-averaged rate independent of $S$
and having a normalized value of $0.014$. The latter regime is triggered by the formation of two chains of plasmoids disrupting
the current sheets with a sudden super-Alfv\'enic growth following a quiescent phase, in agreement with the general theory of the plasmoid instability proposed by
Comisso et al. [{\em Phys. Plasmas} {\bf 23}, 100702 (2016)]. Moreover, the non-monotonic dependence of the plasmoid growth rate with $S$ following
an asymptotically decreasing logarithmic law in the infinite $S$-limit is confirmed.
We also closely compare our results to those obtained during the development of the coalescence instability setup in order to
assess the generality of the mechanism. Finally, we briefly discuss the relevance of our results to explain the flaring activity in solar corona 
and internal disruptions in tokamaks.

\end{abstract}


\section{Motivation}

Magnetic reconnection is believed to be the underlying mechanism that explains explosive events
observed in many magnetically dominated plasmas. This is for example the case for flares in the solar corona, or
sawtooth crashes in tokamaks. It is a process of topological rearrangement of magnetic field lines that
can convert a part of the magnetic energy into kinetic energy and heat \citep {pri00}.
However, the timescales involved
in classical two-dimensional (2D) reconnection models within the macroscopic magnetohydrodynamic (MHD) regime are too slow to match the observations or experiments.
Indeed, the reconnection rate predicted by Sweet-Parker (SP) model which scales like $S^{-1/2}$ ($S$ being the
Lundquist number defined as $S = L V_A / \eta$,  where $L$ is the half-length of the current sheet, $V_A$ is the Alfv\'en speed based
on the magnetic field amplitude in the upstream current layer, and $\eta$ is the resistivity), is too low by a few (or even many) orders of magnitude
for the relevant Lundquist numbers \citep {swe58, par57}.
For example, for typical parameters representative of the solar corona, $S$ is of order $10^{12}$, leading to a normalized reconnection rate of order
$10^{-6}$ much lower than the value of $10^{-2} -10^{-1}$ required to match the observations. Furthermore, SP theory assumes a
steady-state process that cannot explain the impulsive (thus even faster) onset phase preceding the main one.

However, it has been realized in the last decade that, even in a magnetofluid approach, a new solution with a rate that is (possibly) fast
enough and almost independent on $S$ can be obtained, provided that $S$ is higher than a critical value of order $10^4$.
This new regime is supported by the formation of plasmoid chains disrupting the current sheet in which they are born, as
obtained in many numerical experiments \citep {sam09, bha09, hua10}.
More precisely, these plasmoids are small magnetic islands due to tearing-type resistive instabilities, constantly forming, moving,
eventually coalescing, and finally being ejected through the outflow boundaries. At a given time, the system appears as an aligned layer structure of plasmoids of
different sizes, and can be regarded as a statistical steady state with a time-averaged reconnection rate that is nearly (or
exactly) independent of the dissipation parameters \citep {uzd10, lou12}.
The linear modal theory of plasmoid instability is based on a preformed static
(i.e. reconnection flows effects are neglected) unstable SP current sheet with a half-width $a \simeq L S^{-1/2}$ \citep {lou07}.
Among the spectrum of many unstable modes (as $k a  \leq 1$ is required if we assume a Harris-type current layer profile having an hyperbolic tangent magnetic field reversal),
the linearly dominant wavenumber $k_p$ follows $k_p L  \simeq 1.4$ $(1 + P_m)^{-3/16} S^{3/8}$ (where $P_m = \nu/\eta$ is the magnetic Prandtl number, i.e. the ratio of
viscosity coefficient $\nu$ and resistivity one $\eta$) with a corresponding maximum linear growth rate $\gamma_p$ scaling as $\gamma_p  \tau_A  \simeq 0.62$  $(1 + P_m)^{-5/8} S^{1/4}$,
where $\tau_A = L/V_A$ is the Alfv\'en time based on the current sheet half-length \citep {com16, hua19}.

Beyond these above well admitted results and despite many published papers on the subject, there is no clear consensus
on a theoretical view for the plasmoids-reconnection regime including the onset phase.

The paradoxal result of infinite linear growth rate (see scaling law just above) in an ideal MHD plasma (i.e. infinite $S$) being incompatible with the frozen-in condition that makes
reconnection impossible, an issue has been proposed by considering unstable current layers having a critical aspect ratio $L/a \simeq S^{ \alpha}$, that is
smaller than SP value in the high $S$ limit as $0.25 < \alpha < 0.5$ \citep {puc14}. In this way, the linear growth rate becomes Alfv\'enic and independent of $S$.
The value of the exponent $\alpha$ depends on the current profile \citep {puc18}.
For example,  $\alpha=1/3$ is found for the standard Harris current profile, leading to 
$\gamma_p \tau_A  \simeq 0.62$ (using zero viscosity)
with the corresponding linearly dominant wavenumber $k_p$ following
the relation $k_p L  \simeq 1.4$ $S^{1/6}$. These results have been confirmed by numerical simulations of preformed static current layers having the correct aspect ratio value, and
seem to remain true when extended to macroscopic current sheets (of fixed length) that are artificially forced to collapse 
asymptotically towards $a/L \sim S^{-1/3}$ and $a/L \sim S^{-1/2}$
on a time scale of order of $\tau_A$ \citep {ten15, ten16}.  

On the other hand, a second theoretical issue has been proposed by \citet {com16b, com17}
by investigating the plasmoid instability in a dynamically evolving (exponentially shrinking in time and reaching asymptotically a SP aspect ratio)
current sheet. Without any assumption on the critical current sheet aspect ratio (for disruption onset), they employ a principle of least time to 
derive it as well as the corresponding dominant mode and associated growth rate. The main difference compared to the approach proposed in the first issue,
is that the dominant mode is not necessarily the linearly fastest one (obtained from a static stability study), but the mode that is able to emerge first at the
end of the linear phase. In this way, new scalings that are not simple $S$-power laws
are obtained. For example, the dominant mode growth rate is predicted to follow
a transition between the previous scaling as $S^{1/4}$ (for $S$ close to $S_c$) and an asymptotic (for infinitely high $S$ values) new scaling following a decreasing logarithmic
dependence (see Equation 19 in \citet {com16b} and Equation 32 in \citet {com17}).  
The growth rate can in principle easily attain super-Alfv\'enic values  $\gamma_p  \tau_A  \sim 10 - 100$, 
while remaining finite in the infinite $S$ limit. The precise value of the growth rate and of the corresponding wavenumber also depend on other parameters than $S$,
that are the characteristic time scale of the current sheet formation, the thinning process, the magnetic Prandtl number, and the noise of the system.

This second issue seems to be partly supported by recent 2D numerical MHD simulations, where the coalescence instability between
two parallel currents is chosen as
the initial setup providing the thinning process to form the current sheet \citep {hua17}. Indeed, a scaling transition
is effectively observed, and maximum growth rates with $\gamma_p \tau_A \simeq 10-20$ 
are obtained that are substantially smaller than values predicted by the theoretical model.
The remaining differences between the simulations and the analytical model of
\citet {com16b, com17} are explained by taking into account the effects of the reconnection outflow in a phenomenological model \citep {hua19}.
Conversely, as the first theoretical model proposed by \citet {puc14} predicts constant and smaller 
 growth rates, i.e. with $\gamma_p \tau_A  \sim 1$, it consequently 
seems to fail to explain these numerical simulations. 

Consequently, this is important to use other configurations in order to address the generality of the reconnection mechanism based
on this plasmoid-unstable regime. 
We have thus chosen to consider a different setup, that is the tilt instability between two repelling
antiparallel currents \citep {ri90}. Curiously, this latter configuration has been used to test ideal MHD codes \citep {lan07},
or more recently to study the interaction with kink instability and resulting particles acceleration \citep {kep14, rip17},
but not to study magnetic reconnection associated to the plasmoids formation. 
Note that, a preliminary study mainly devoted to demonstrate the ability of our specifically designed MHD code, FINMHD, has been recently published 
 \citep {bat19}. In the present work, we mainly focus on the onset phase leading to the disruption of the current sheets by the formation of many
plasmoids. The ensuing statistical state with a fast reconnection rate is also addressed more superficially, as a full assessment
of this state is beyond the scope of the present paper and is left to a future work.
The outline of the paper is as follows. In Section 2, we present the MHD code and the initial setup for tilt instability.
Section 3 is devoted to the presentation of the results. In section 4, our results are compared to those obtained using coalescence
setup and to those predicted by the two theoretical models cited above. Finally, we conclude in section 5.

\section{The MHD code and initial setup}

\subsection{FINMHD equations}
For FINMD, a set of reduced MHD equations has been chosen corresponding to a 2D incompressible model. However,
instead of taking the usual formulation with vorticity and magnetic flux functions for the main variables, another choice
using current-vorticity ($J-\omega$) variables is preferred because of its more symmetric formulation, facilitating the numerical
matrix calculus. The latter choice also cures numerical difficulty due to the numerical
treatment of a third order spatial derivative term \citep{phi07}. To summarize, the following set of equations is (see also \citet{bat19} for more details),
\begin{equation}  
      \frac{\partial \omega}{\partial t} + (\bm{V}\cdot\bm{\nabla})\omega = (\bm{B}\cdot\bm{\nabla})J + \nu \bm{\nabla}^2 \omega ,
\end{equation}
\begin{equation}      
        \frac{\partial J }{\partial t} + (\bm{V}\cdot\bm{\nabla})J =  (\bm{B}\cdot\bm{\nabla})\omega + \eta \bm{\nabla}^2 J +  g(\phi,\psi) ,
\end{equation}
\begin{equation}                 
     \bm{\nabla}^2\phi = - \omega ,
 \end{equation}
\begin{equation}                        
     \bm{\nabla}^2\psi = - J ,  
\end{equation}
with $g(\phi,\psi)=2 \left[\frac{\partial^2 \phi}{\partial x\partial y}\left(\frac{\partial^2 \psi}{\partial x^2} - \frac{\partial^2 \psi}{\partial y^2}\right) - \frac{\partial^2 \psi}{\partial x\partial y}\left(\frac{\partial^2 \phi}{\partial x^2} - \frac{\partial^2 \phi}{\partial y^2}\right)\right]$.
As usual, we have introduced the two stream functions, $\phi (x, y)$ and $\psi (x, y)$, from the fluid velocity $\bm{V} = {\nabla} \phi \wedge \bm{e_z}$ and magnetic field $\bm{B} = {\nabla} \psi \wedge \bm{e_z}$ ($\bm{e_z}$
being the unit vector perpendicular to the $xOy$ simulation plane).
$J$ and vorticity $\omega$ are the $z$ components of the current density and vorticity vectors, as $\bm{J} = \nabla \wedge \bm{B}$ and $\bm{ \omega} = \nabla \wedge \bm{V}$ respectively (with units using $\mu_0 = 1$).
Note that we consider the resistive diffusion via the $\eta \bm{\nabla}^2 J $ term ($\eta$ being assumed uniform for simplicity), and also a viscous term
$\nu \bm{\nabla}^2 \omega$ in a similar way (with $\nu$ being the viscosity parameter also assumed uniform).
The above definitions results from the choice $\psi \equiv A_z$, where $A_z$ is the $z$ component of the potentiel vector $\bm{A}$ (as $\bm{B} = \nabla \wedge \bm{A}$). This choice
is the one used in \citet{ng07} or in \citet{bat16},
and different from the one used by \citet{lan07} where the choice $\psi \equiv - A_z$ is done.
In the latter case, the two Poisson equations (i.e. Equations 3-4) involve an opposite sign in the right hand sides. Note that thermal pressure gradient is naturally absent from our set of equations. 
Note also that, an advantage of the above formulation over a standard one using the velocity and magnetic field vectors ($\bm{V}, \bm{B}$)  as
the main variables, is the divergence-free property naturally ensured for these two vectors. 

\begin{figure}
\centering
 \includegraphics[scale=0.17]{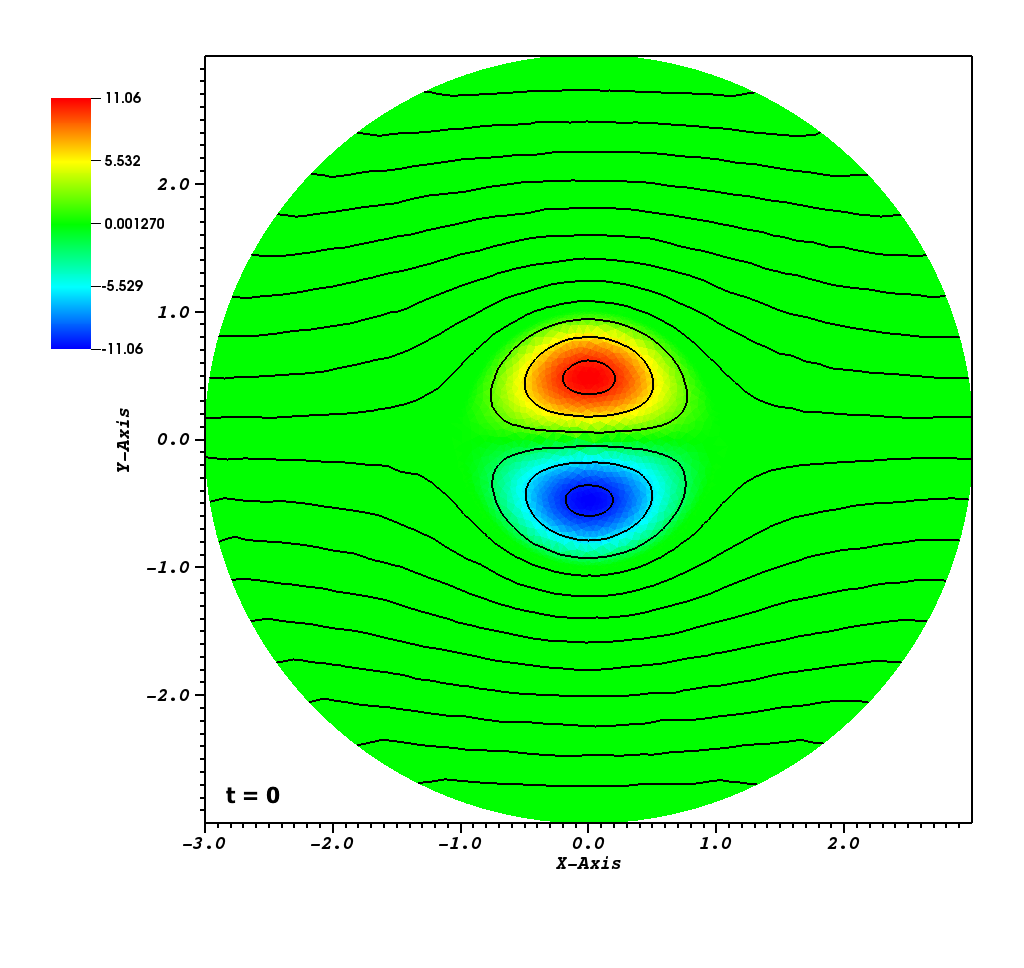}
 \includegraphics[scale=0.17]{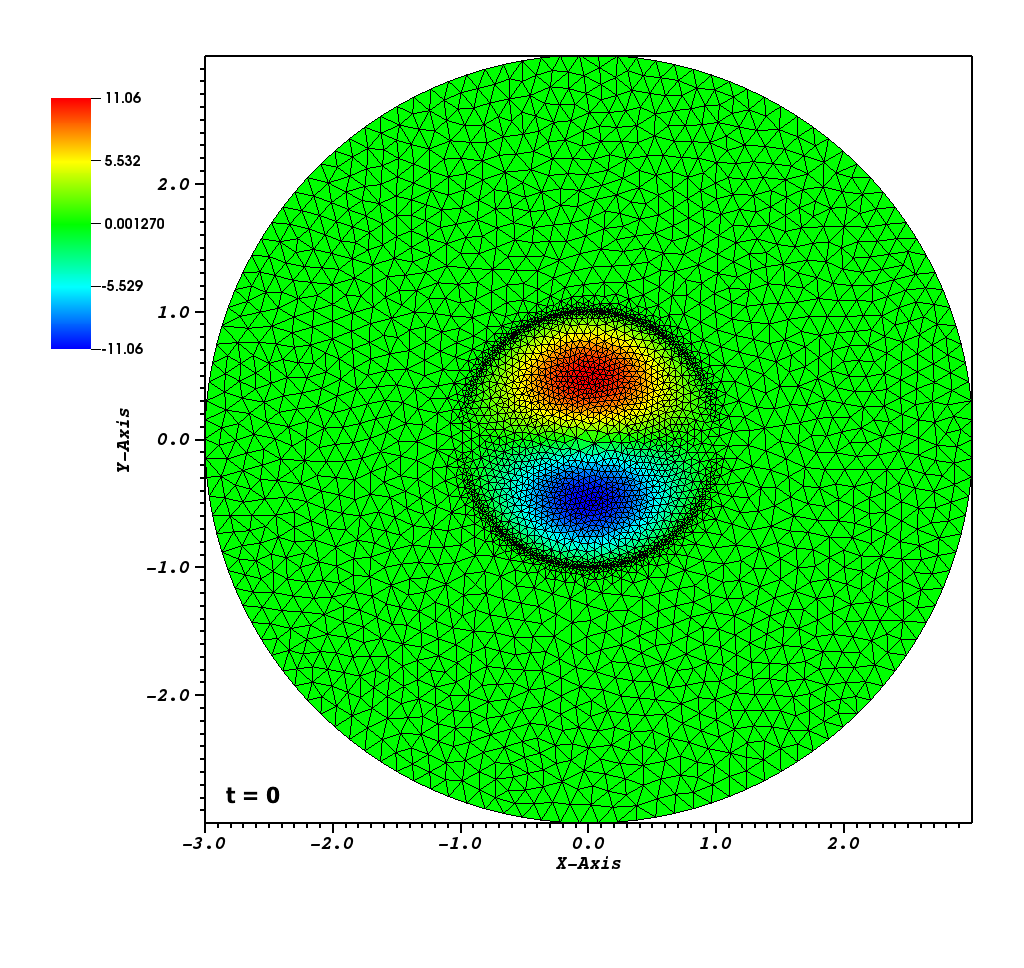}
  \caption{Initial configuration for the current density structure (colored contour map) overlaid with magnetic field lines (left panel), and overlaid 
  with the initial grid using the density density to adapt the mesh (right panel). A moderately high value for the maximum edge size
  of $h_m = 0.05$ is imposed for this case using isogeometric triangles.
  }
\label{fig1}
\end{figure}

\begin{figure}
\centering
 \includegraphics[scale=0.17]{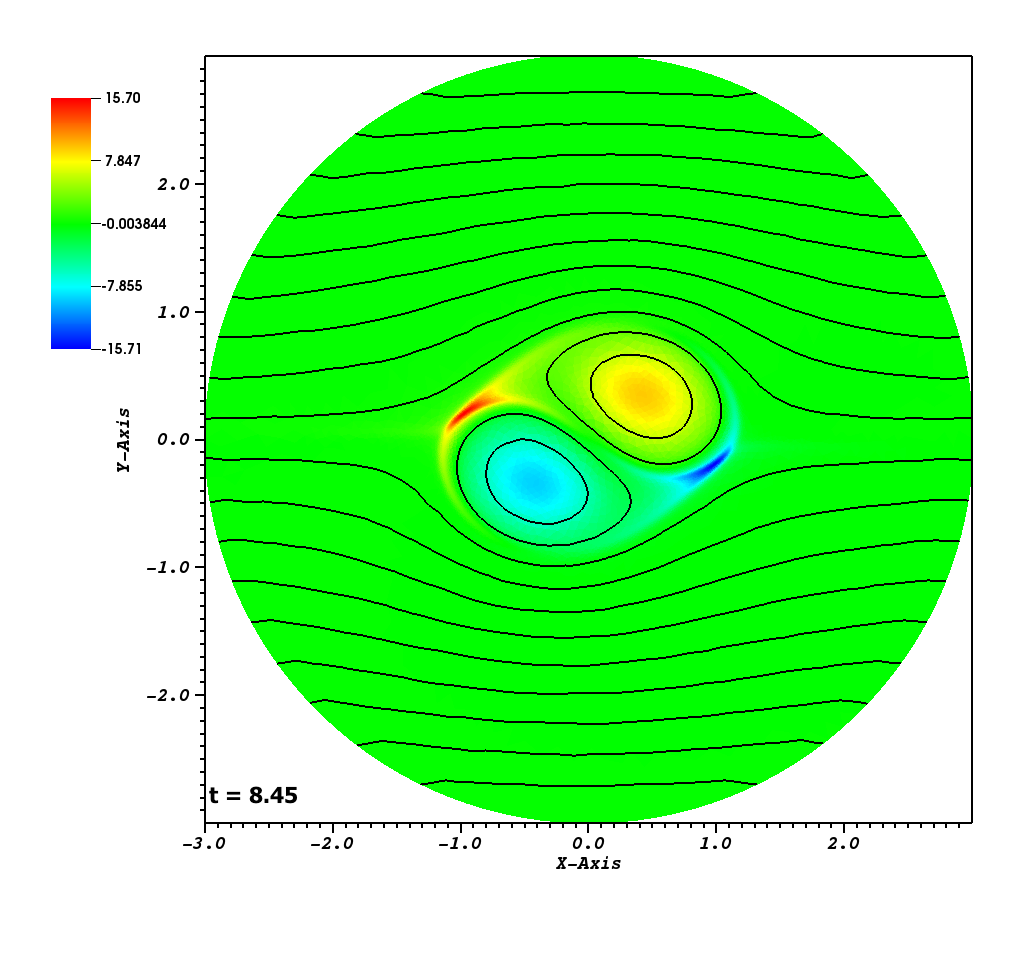}
 \includegraphics[scale=0.17]{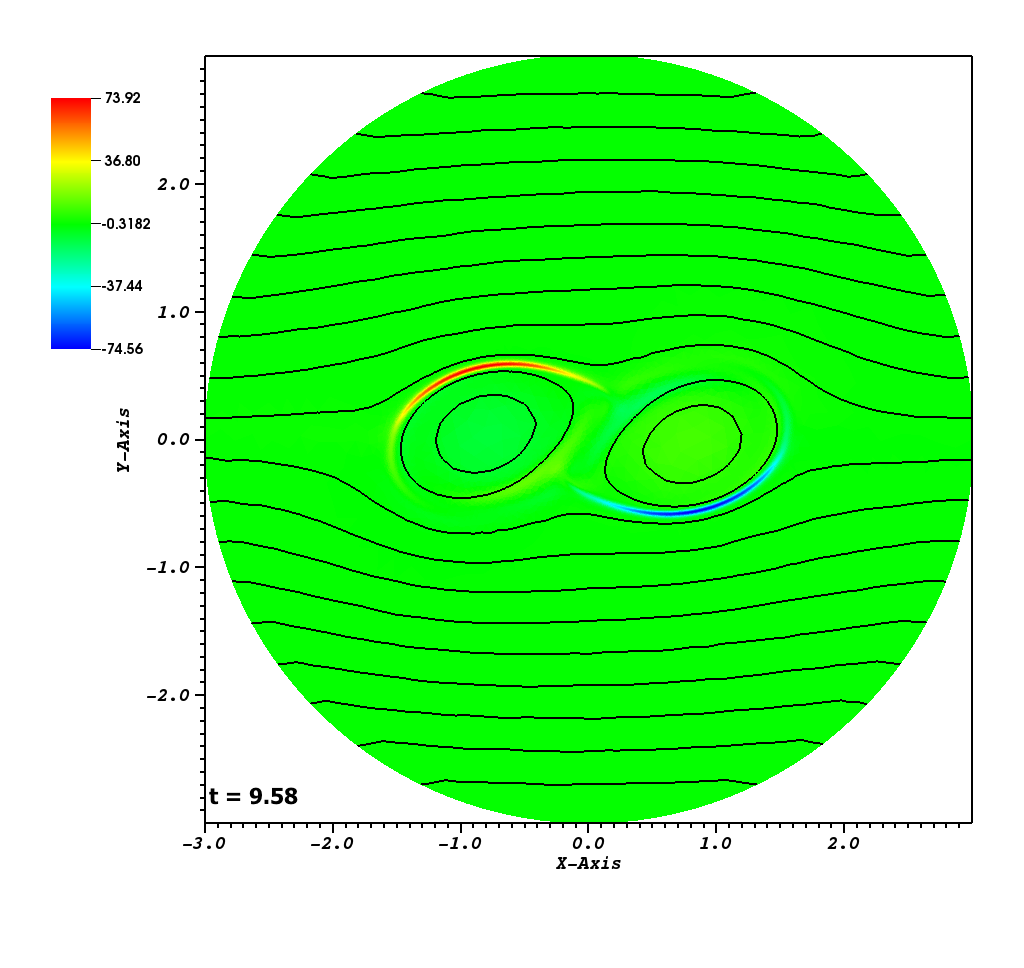}
  \caption{Same as Figure 1 for two times ($t = 8.45$ $t_A$ and $t = 9.58$ $t_A$) during the development of the tilt instability, using $S^* = 10^3$ (i.e. $\eta = \nu = 1 \times 10^{-3}$). }
\label{fig2}
\end{figure}

\subsection{FINMHD numerical method}

Simulating the mechanism of magnetic reconnection in the high Lundquist number regime requires the use of
particularly well adapted methods. Conventional codes generally lack some convergence properties
to follow the associated complicated time dependent bursty dynamics \citep{kep13}. 
Despite the fact that they are not commonly used, finite element techniques allows to treat the early formation of
quasi-singularities \citep{str98, lan07}, and the ensuing magnetic reconnection in an efficient way \citep{bat19}.

FINMHD code is based on a finite element method using triangles with quadratic
basis functions on an unstructured grid. A characteristic-Galerkin scheme is chosen 
in order to discretize in a stable way the Lagrangian derivative $\frac{\partial  }{\partial t} + (\bm{V}\cdot\bm{\nabla}) $ appearing
in the two first equations \citep{bat19}.
Moreover, a highly adaptive (in space and time) scheme is developed in order to follow the rapid
evolution of the solution, using either a first-order time integrator (linearly unconditionally stable) or a second-order one
(subject to a CFL time-step restriction). Typically, a new adapted grid can be computed at each time step, by searching the grid
that renders an estimated error nearly uniform. The finite elements Freefem++ software allows to do this  \citep{hec12}, by using
the Hessian matrix of a given function (taken to be the current density in this study).
The technique used in FINMHD has been tested on challenging tests, involving 
unsteady strongly anisotropic solution for the advection equation, formation of shock structures
for viscous Burgers equation, and magnetic reconnection for the reduced set of MHD equations.
The reader should refer to \citet{bat19} for more details.

\subsection{The initial setup}

The initial magnetic field configuration for tilt instability is a dipole current structure similar to the dipole vortex flow pattern in
fluid dynamics \citep{ri90}.
It consists of two oppositely directed currents embedded in a constant magnetic field (see Figure 1).
Contrary to the coalescence instability based on attracting parallel current structures, the two antiparallel currents in the configuration tend to repel.
The initial equilibrium is thus defined by taking the following magnetic flux distribution,
\begin{equation}
    \psi_e (x, y)=
    \left\{
      \begin{aligned}
        &\left(\frac{1}{r} - r\right)\frac{y}{r} ~~~& & if ~~ r > 1 , \\
        &-\frac{2}{\alpha J_0(\alpha)}J_1(\alpha r)\frac{y}{r} ~~~& & if ~~ r\leq1 .\\
      \end{aligned}
      \right.
  \end{equation}
  
 And the corresponding current density is,
       \begin{equation} 
    J_e (x, y) =
    \left\{
      \begin{aligned}
        &~~~~~~~~~~~~0 ~~~& & if ~~ r > 1 , \\
        &-\frac{2\alpha}{J_0(\alpha)}J_1(\alpha r)\frac{y}{r} ~~~& & if ~~ r\leq1 ,\\
      \end{aligned}
    \right.
\end{equation}

\noindent where 
$r=\sqrt{x^2+y^2}$, and $J_0$ et $J_1$  are Bessel functions of order $0$ and $1$ respectively.
Note also that $\alpha$ is the first (non zero) root of $J_1$, i.e. $\alpha = 3.83170597$.
\medskip
This initial setup is similar to the one used in the previously cited references \citep{ri90},
and rotated with an angle of $\pi/2$ compared to the
equilibrium chosen in the other studies \citep{kep14}.
Note that, the asymptotic (at large $r$) magnetic field strength 
is unity, and thus defines our normalisation. Consequently, our unit time in the following paper, will be defined as the Alfv\'en transit
time across the unit distance (i.e. the initial characteristic length scale of the dipole structure) as $t_A = 1$. The latter time
is slightly different from $\tau_A$ that is based on the half-length of the current sheet and on the upstream magnetic
field magnitude. However, in our simulations we can deduce that $\tau_A \simeq t_A/2$ (see below).
In usual MHD framework using the flow velocity and magnetic variables, force-free equilibria using an additional
vertical (perpendicular to the $x-y$ plane) can be considered \citep{ri90}, or non force-free equilibria
can be also ensured trough a a thermal pressure gradient balancing the Lorentz force \citep{kep14}.
In our incompressible reduced MHD model, as thermal pressure is naturally absent, we are not
concerned by such choice. 

In previous studies using a similar physical setup, a square domain $[-R: R]^2$ was taken with $R$ large enough in order
to have a weak effect on the central dynamics. For example a standard value of $R = 3$ is taken in \citet{bat19}.
In the present work, a choice of using a circular domain with a radius $R = 3$ is done. We have checked that
it does not influence the results compared to the square domain setup. However, this allows the use of a lower number of
finite-element triangles (as the circle area is evidently smaller than the square for the same radius value $R$), and this
also simplifies the numerical boundary treatment as only one boundary instead of four in our finite-element
discretization are needed.

A stability analysis in the reduced MHD approximation using the energy principle has given that the linear
eigenfunction of the tilt mode is a combination of rotation and outward displacement \citep{ri90}. Instead of
imposing such function in order to perturb the initial setup, we have chosen to let the instability develops from
the initial numerical noise. Consequently, an initial zero stream function is assumed $ \phi_e (x, y) = 0$,
with zero initial vorticity $ \omega_e (x, y) = 0$.  The values of our four different variables are also
imposed to be constant in time and equal to their initial values at the boundary $r = R$.

\begin{figure}
\centering
 \includegraphics[scale=0.22]{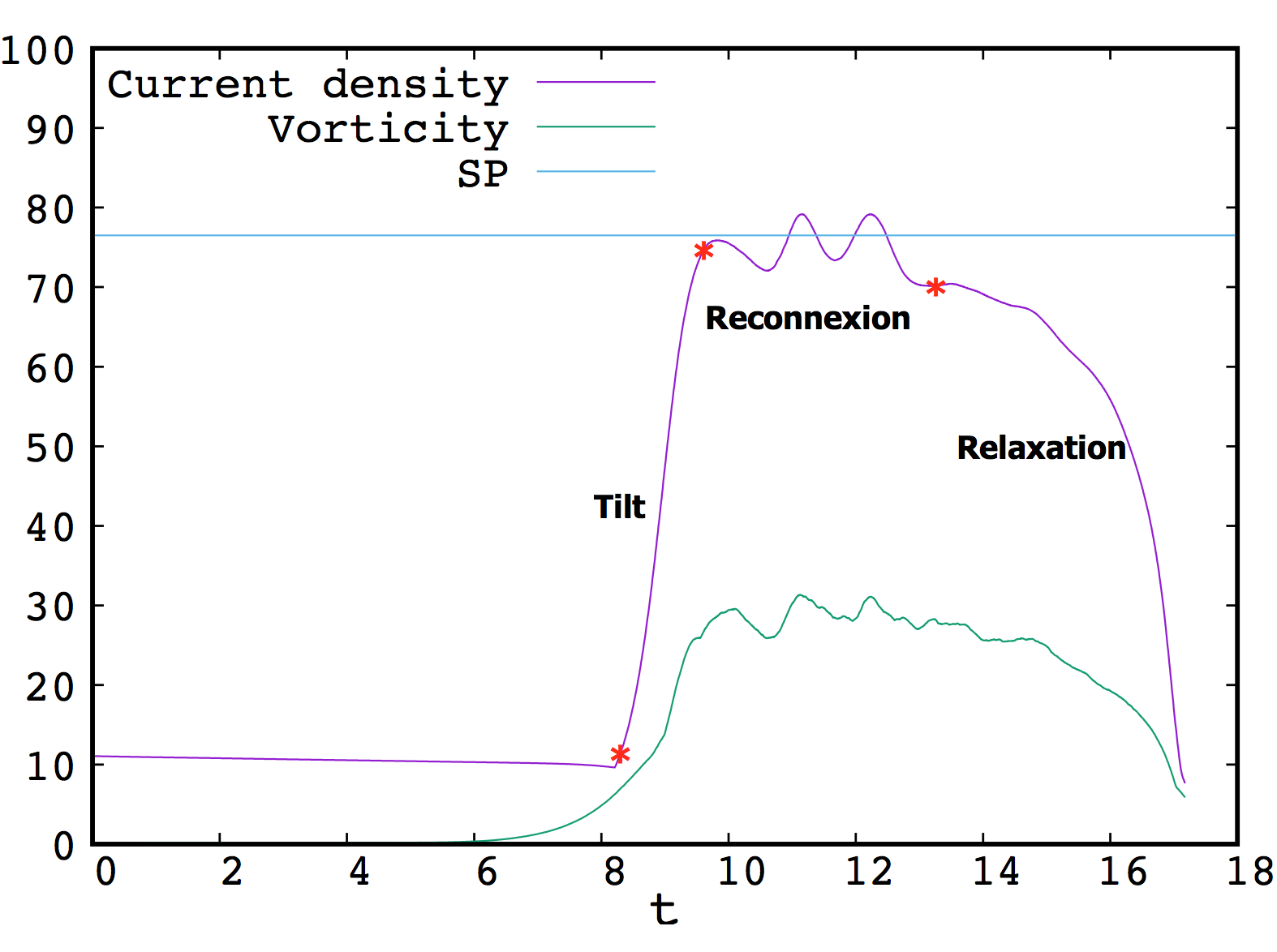}
 \includegraphics[scale=0.22]{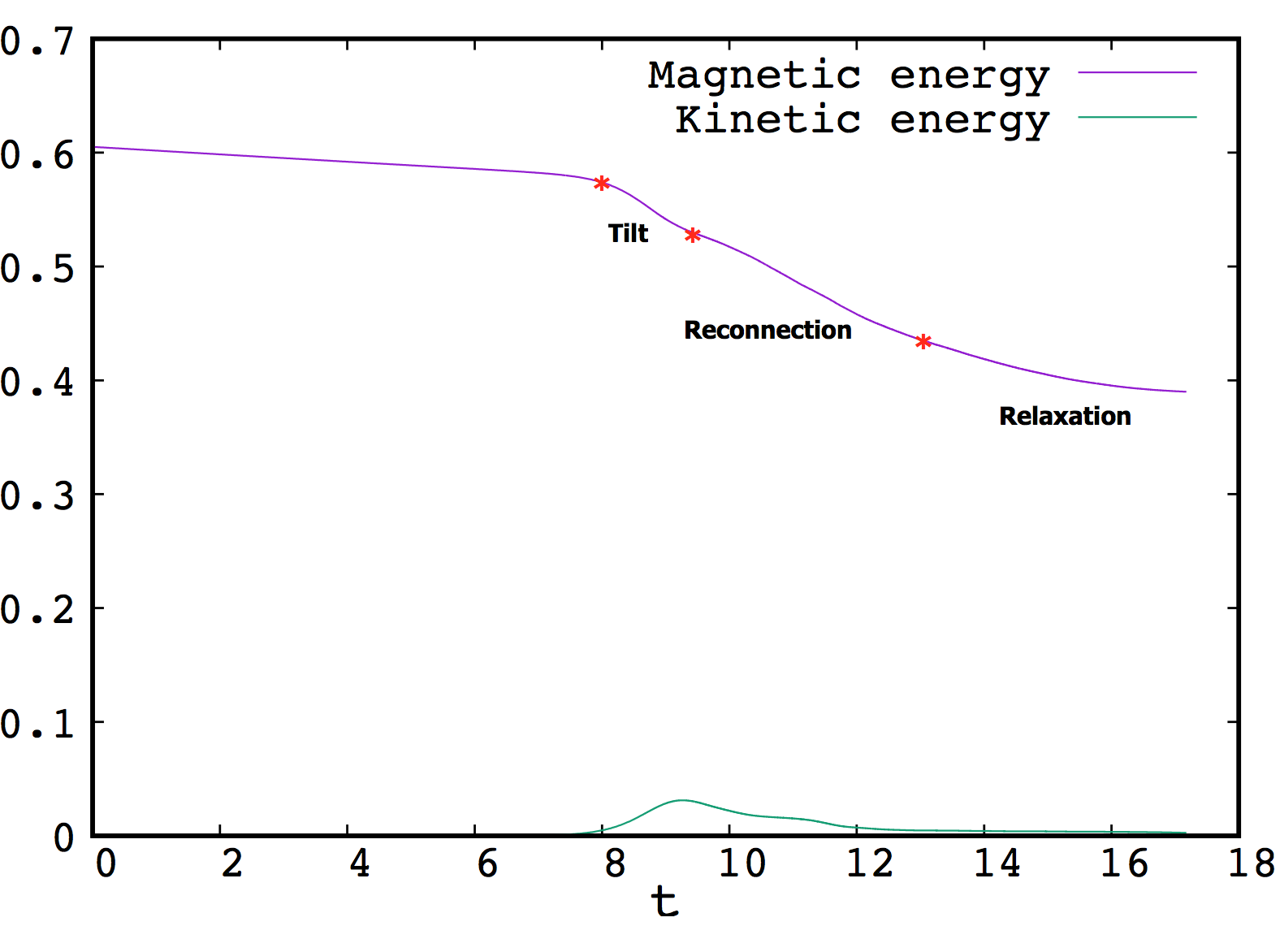}
  \caption{Time history of the maximum current density and maximum vorticity amplitudes for the run using $\eta = \nu = 1 \times 10^{-3}$
  (left pannel). An horizonthal line indicates the average current value (i.e. $77$) during reconnection phase, which also agree with the value
  predicted from the SP theory. Corresponding time history for the magnetic and kinetic energies in our normalized units (right panel).}
\label{fig3}
\end{figure}

\begin{figure}
  \centerline{\includegraphics [width=10cm]{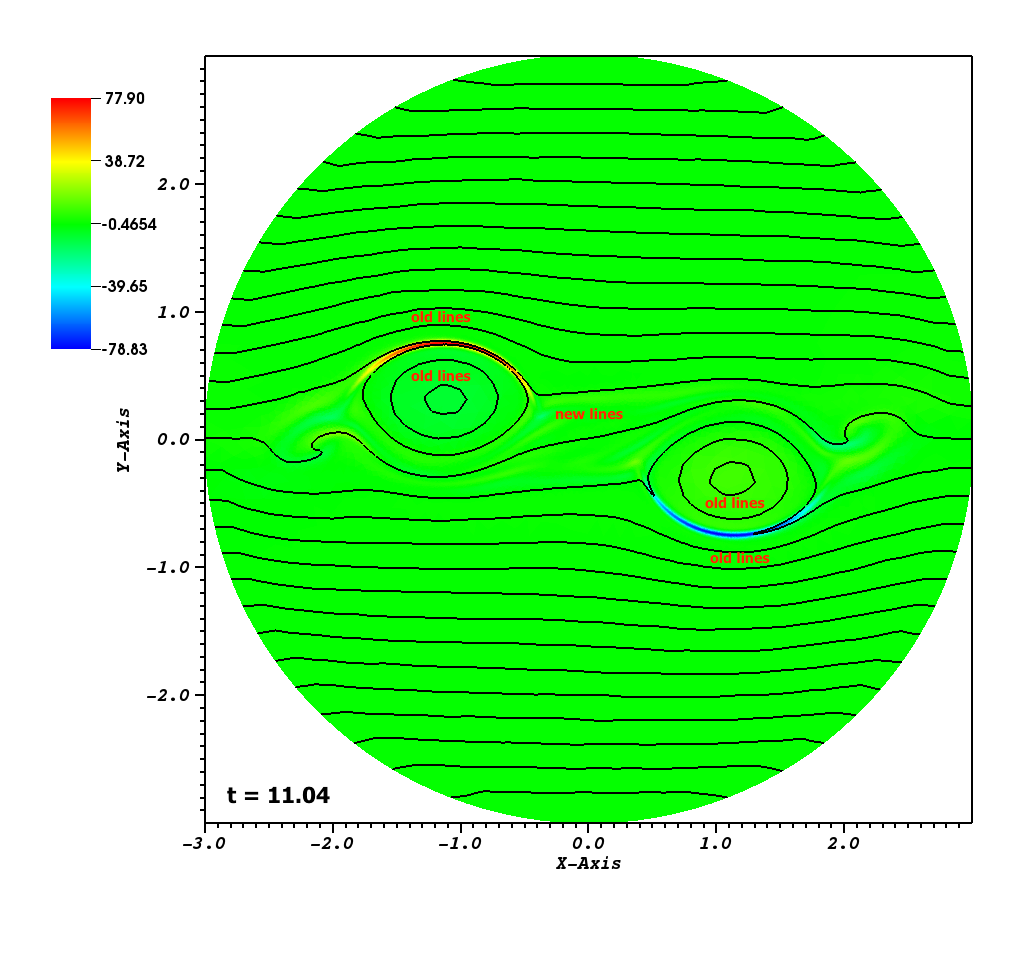}}
  \caption{Same as Figures 1-2 for a later time ($ t = 11.04$ $t_A$) showing the reconnection process with a few magnetic field lines.}
\label{fig4}
\end{figure}

\section{Results}

\subsection{Initial development of the tilt instability and SP reconnection regime}

The initial current structure with a few magnetic field lines are reported in Figure 1 (left panel), with the initial mesh that is adapted using the Hessian
matrix of the current density distribution (right panel). This arbitrary choice of current density is justified by the fact that, first the initial dynamics of the tilt instability
is driven by the current distribution (i.e. this is an ideal current driven MHD instability), and second the ensuing magnetic process is controlled by the structure
of the current layers. Numerical parameters defining the maximum edge size ($h_m$) and the anisotropy of the triangular mesh
have been also adapted from case
to case. Indeed, the value of $h_m$ must be decreased as small resistivity/viscosity coefficients are employed in order to capture the small scale turbulence,
leading however to a higher number of triangles and increase of time computing. The efficiency of the latter procedure is demonstrated in \citet{bat19}.

First, we focus on moderately low values of the resistivity and viscosity, and for simplicity we also assume a fixed Prandtl number,
$P_r = \nu / \eta = 1$ in the whole paper. The early time evolution of the system corresponding to the tilt instability is well documented (see the previously cited studies).
It corresponds to the linear stability analysis, where the pair of oppositely directed currents tend to repel one another giving rise to a rotation.
This is illustrated in Figure 2 at two times during the instability development, for a case employing an inverse resistivity value $S^* = 1/\eta = 10^3$. 
Note that, the sense of rotation (clockwise in Figure 2) is not predetermined and depends only on the numerical noise.
This rotation causes two new regions of enhanced current density (having opposite sign) at the leading edges of the vortices because of an associated outward
component of the linear displacement, taking the form of two bananas. These two regions are the relevant forming current layers of our study.

The time history of the maximum current density and vorticity amplitudes (taken over the whole domain) are shown to increase exponentially in time,
and to rapidly dominate the equilibrium values (approximately $10$ for the current density at $t  \simeq 8$ $t_A$), as illustrated in Figure 3 (left panel).
Note that, there is a small time delay between the current and vorticity increase. We have checked that, the associated linear dynamics follow
the expected exponential time increase like $e^{2.6 t}$ and $e^{1.4 t}$ for the current density and vorticity respectively in
agreement with stability theory (see \citet{bat19} and references therein).
During the linear and ensuing nonlinear phases of the tilt development, the kinetic energy (see right panel of Figure 3) increases in
correspondence with a small decrease of the magnetic energy (the sum of both being conserved before reconnection takes place).
The reconnection is triggered just before the saturation observed in current density (second asterisk in Figure 3). 
Subsequently, a steady-state reconnection regime is obtained with nearly constant current density/vorticity structures.
The oscillations (in the maximum current density and vorticity) around average values are due to the sloshing phenomenon,
as described in the coalescence problem between two magnetic islands \citep{kno06}, because of the magnetic pressure buildup effect in
thin current sheets.  An average value of $77$ is evaluated for the maximum current density during magnetic reconnection, in agreement
with SP theory (see below). The process of
magnetic reconnection between each current channel (circular magnetic field lines) and the background magnetic field (open straight field lines)
is clearly visible in Figure 4, where the new reconnected field lines are also plotted.
Finally, a last phase that is a relaxation towards a new state free of closed circular magnetic field lines is obtained. The latter phase
is shown to begin (spotted by the last asterisk in Figure 3) when roughly $80-90$ per-cent of the free magnetic energy is released.
The detailed structure of the current sheets structure, the adapted mesh, and corresponding magnetic field lines during reconnection are shown in
Figure 5. Indeed, one can clearly see in right panel of Figure 5, the few tens of elements covering the width of the current layer,
thus illustrating the efficiency of our dynamically adapting mesh procedure.

\begin{figure}
\centering
 \includegraphics[scale=0.17]{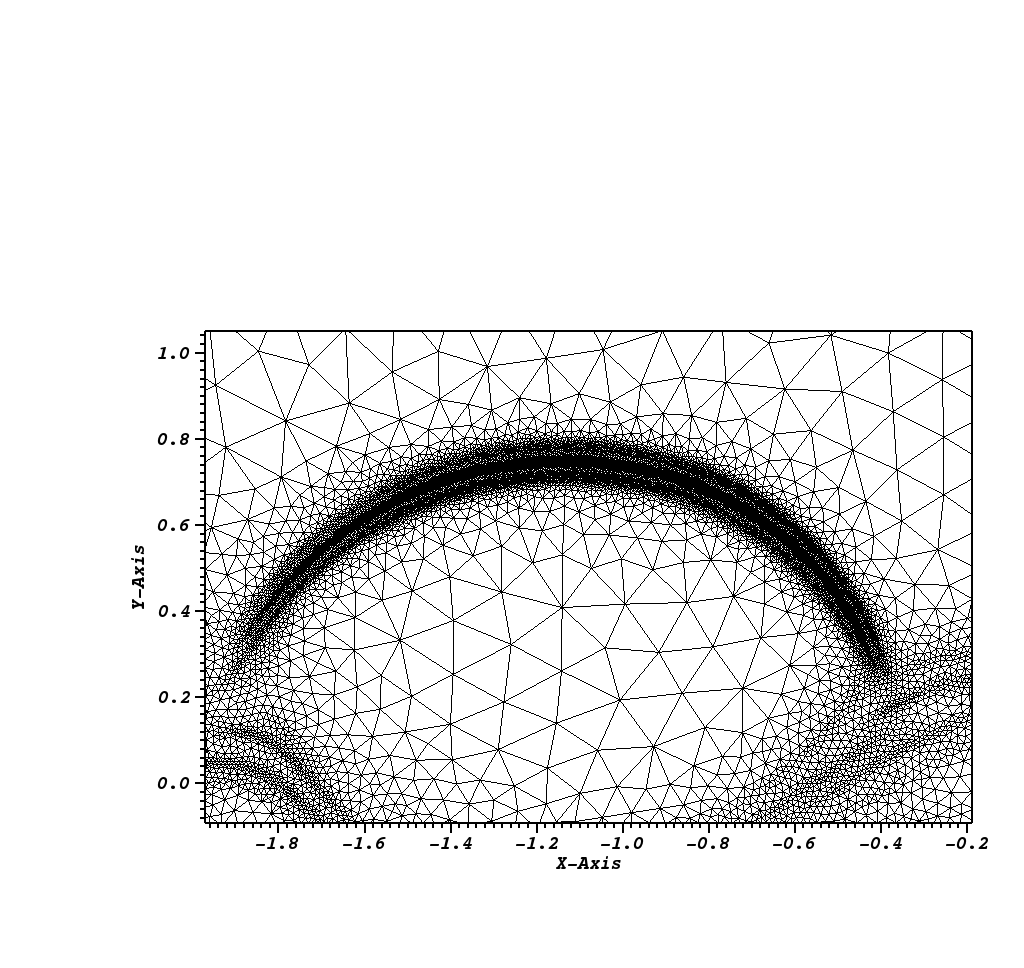}
 \includegraphics[scale=0.17]{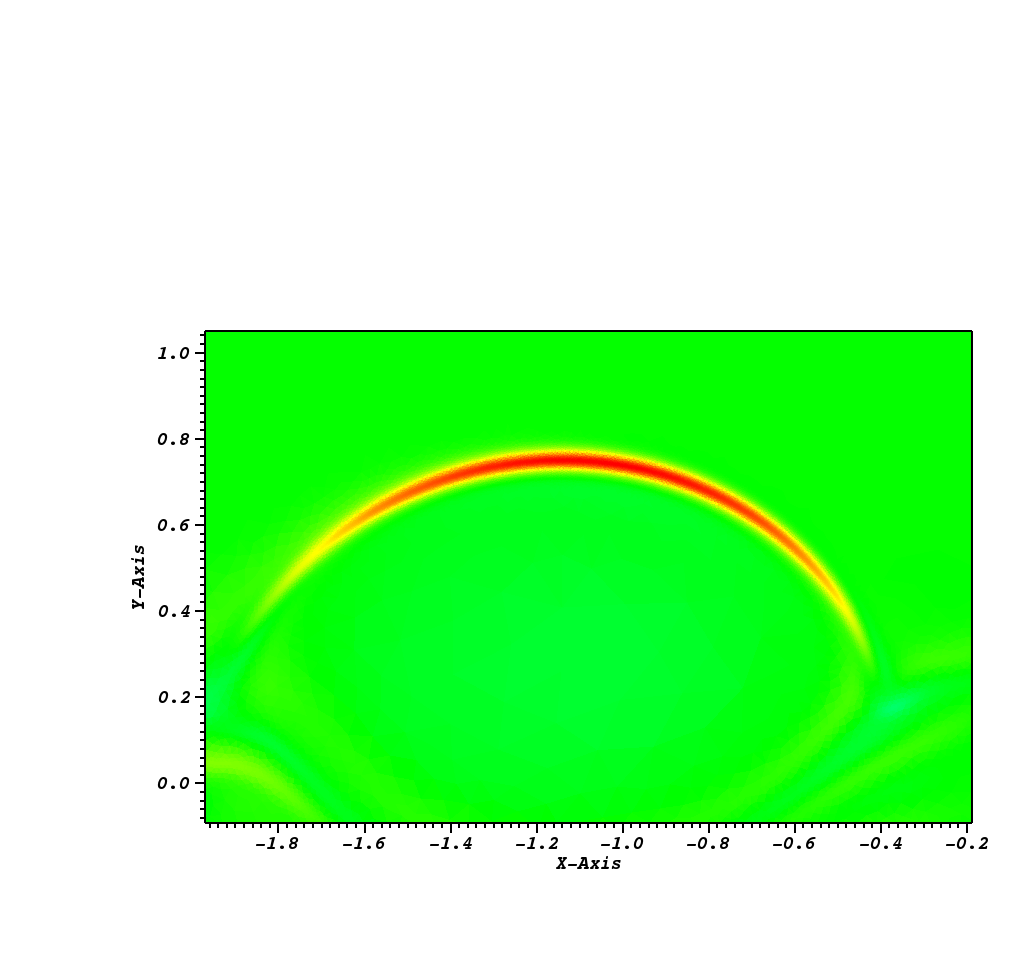}
  \caption{Zoom-in of one current sheet taken from previous figure (left panel), and corresponding adapted mesh (right panel).}
\label{fig5}
\end{figure}

\begin{figure}
\centering
 \includegraphics[scale=0.57]{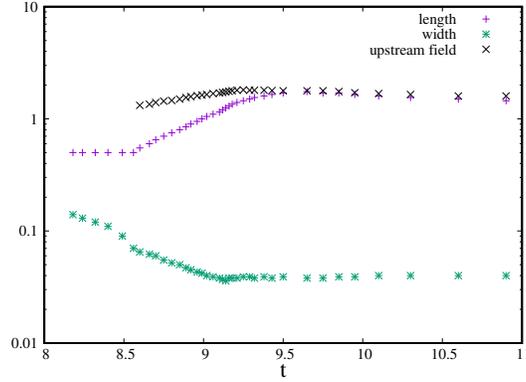}
  \caption{Time history of the characteristics of the forming current sheet (length $2L$, width $2a$, and upstream magnetic field amplitude $B_u$)
  for the run using $\eta = \nu = 1 \times 10^{-3}$ (i.e. $S^* = 10^3$).}
\label{fig6}
\end{figure}

In order to evaluate the local Lundquist number at saturation, $S = L V_A/\eta$, where $L$ is the half-length
of each current layer and $V_A$ is the Alfv\'en velocity based on the upstream magnetic field magnitude $B_u$ measured front of the layer,
we have followed the time history of the length ($2L$), the width ($2a$), and the upstream magnetic field amplitude ($B_u$).
More precisely, the length and width are deduced by evaluating the locations where the current density is decreased
by a factor of two compared to the maximum current density (i.e. obtained at the centre of the current layer). The value of $B_u$
(see Figure 8) is the magnetic field measured in the current region with closed circular field lines, and it is slightly larger than the value measured in the
region opposite to the current sheet with open nearly straight field lines because of the asymmetry.
The results that are plotted in Figure 6 for $S^* = 10^3$, show that the three parameters are varying in time during the
current sheet formation, and not only the width as assumed for example in theoretical models \citep{com16b, com17}. Second, a value of $S = 1500$ is
deduced during reconnection phase, as $B_u \simeq 1.8$ and $L \simeq 0.85$ for this run. 

Finally, we have compared the time evolution of the system in four different runs using inverse resistivity values $S^* = 500, 666, 1000,$ and $1428$.
The results obtained for the maximum current density and vorticity are plotted in Figure 7. The respective corresponding $S$ values have been estimated to be
$S = 625, 890, 1500,$ and $2100$, and remain substantially lower than the critical value $S_c  \simeq 10^4$ required for plasmoid formation.
Note that The values of $L$ and $V_A$ used to deduce the above $S$ values, slightly increase when the resistivity is decreased for $ \eta \simgt 0.0005$,
and they become approximately constant for smaller resistivity values. 
The characteristic reconnection time $\tau_r$ can be determined in different equivalent ways. The simplest one is to take the whole duration of the process,
and it corresponds to the elapsed time between the saturation (first peak) and the
final time when the current density in Figure 3 returns to a very small value corresponding to a new state
(free of closed magnetic field lines). A second way is to measure the level of maximum current amplitude $J_{max}$  during reconnection (horizontal
line in Figure 3), and to multiply it by the resistivity, as $\eta J_{max}$ is a measure of the reconnection rate (inverse of $\tau_r$) for a steady-state
process. We have checked that the reconnection rate for $S  \simlt 10^4$ closely follows a SP scaling, as $J_{max} \simeq 2 \times S^{1/2}$ 
(see also Figure 14), in close agreement with theoretical prediction of $J_{max} = B_u/a \simeq B_u S^{1/2} (1 + P_r )^{-1/4} /L$ as
the SP aspect ratio itself is predicted to be $L/a \simeq S^{1/2} (1 + P_r )^{-1/4}$ \citep{com16,par84}. In a similar way, the maximum vorticity
is shown to follow a SP scaling, as $\Omega_{max} \simeq 0.75 \times S^{1/2}$, that is a factor of two smaller than predicted by the theoretical
formula $\Omega_{max} = V_o/a \simeq  S^{1/2} (1 + P_r )^{-3/4} V_A/L \simeq  1.4 \times S^{1/2}$ ($V_{out}$ being the outflow SP velocity, $V_{out} \simeq (1 + P_m)^{-1/2} V_A$).
This factor of two clearly comes from the asymmetric geometry of the current sheet, as schematized in Figure 8. $\Omega_{max} = V_{out}/(2 a)$ is consequently a better
definition formula four our problem.

\begin{figure}
\centering
 \includegraphics[scale=0.47]{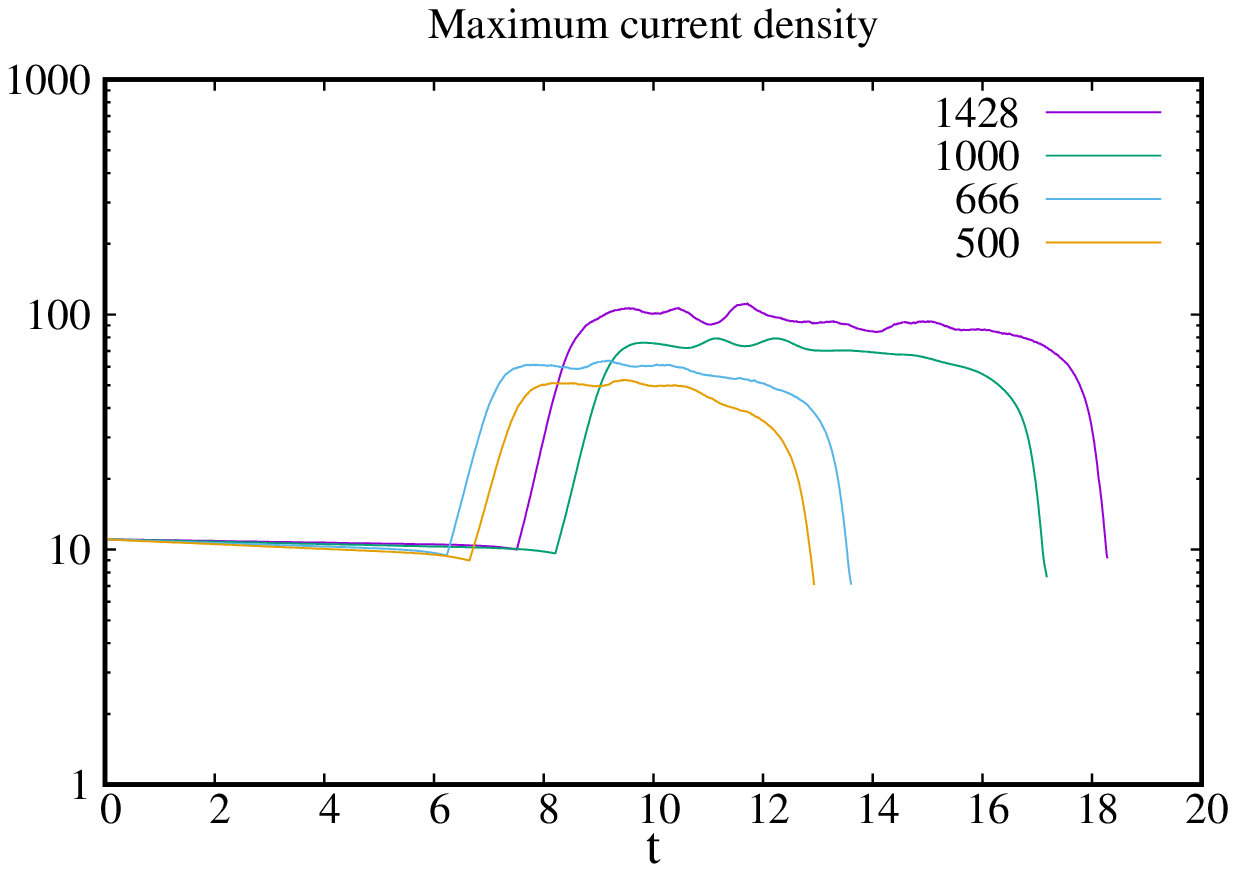}
 \includegraphics[scale=0.47]{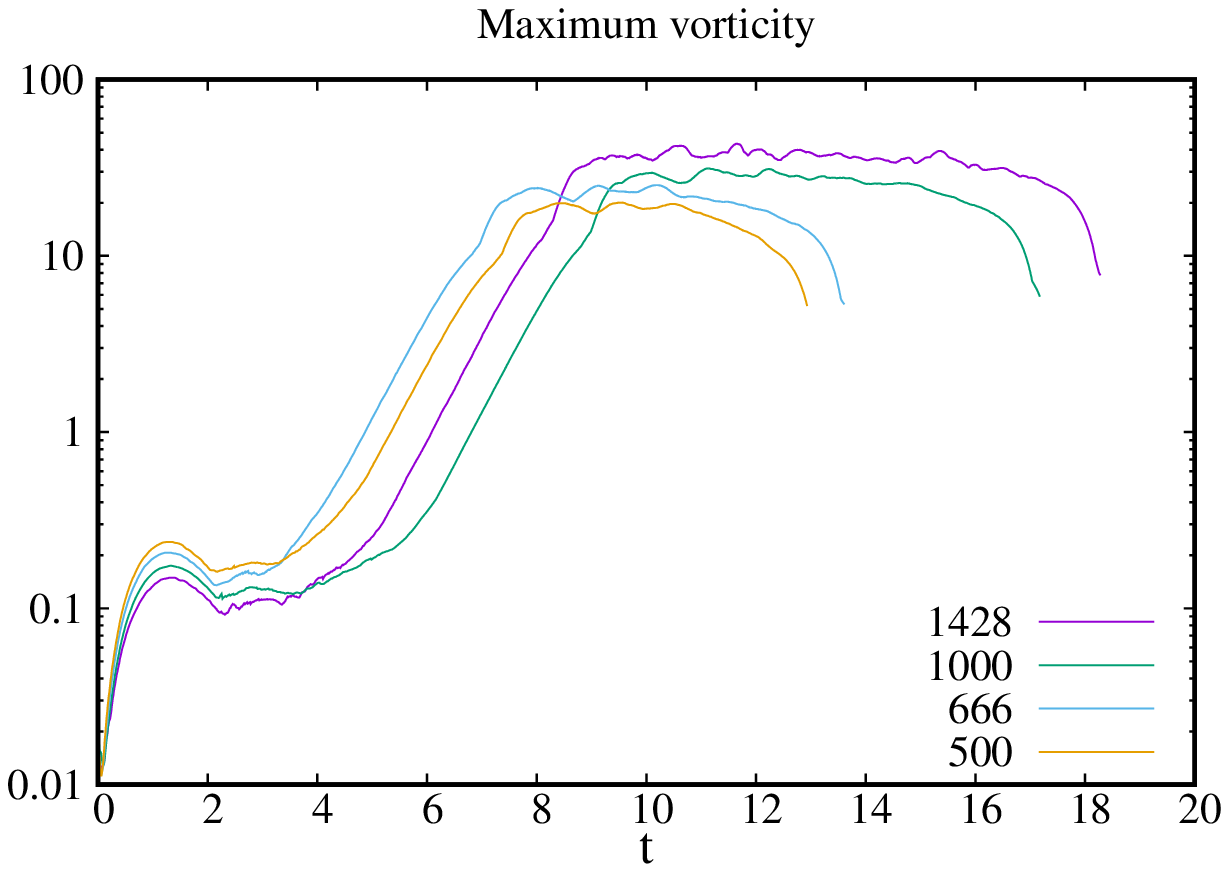}
  \caption{Time history of the maximum current density (left panel), and maximum vorticity (right panel) amplitudes for $4$
  runs employing inverse resistivity values $S^*$ (between $500$ and $1428$).}
\label{fig7}
\end{figure}

\begin{figure}
\centering
 \includegraphics[scale=0.27]{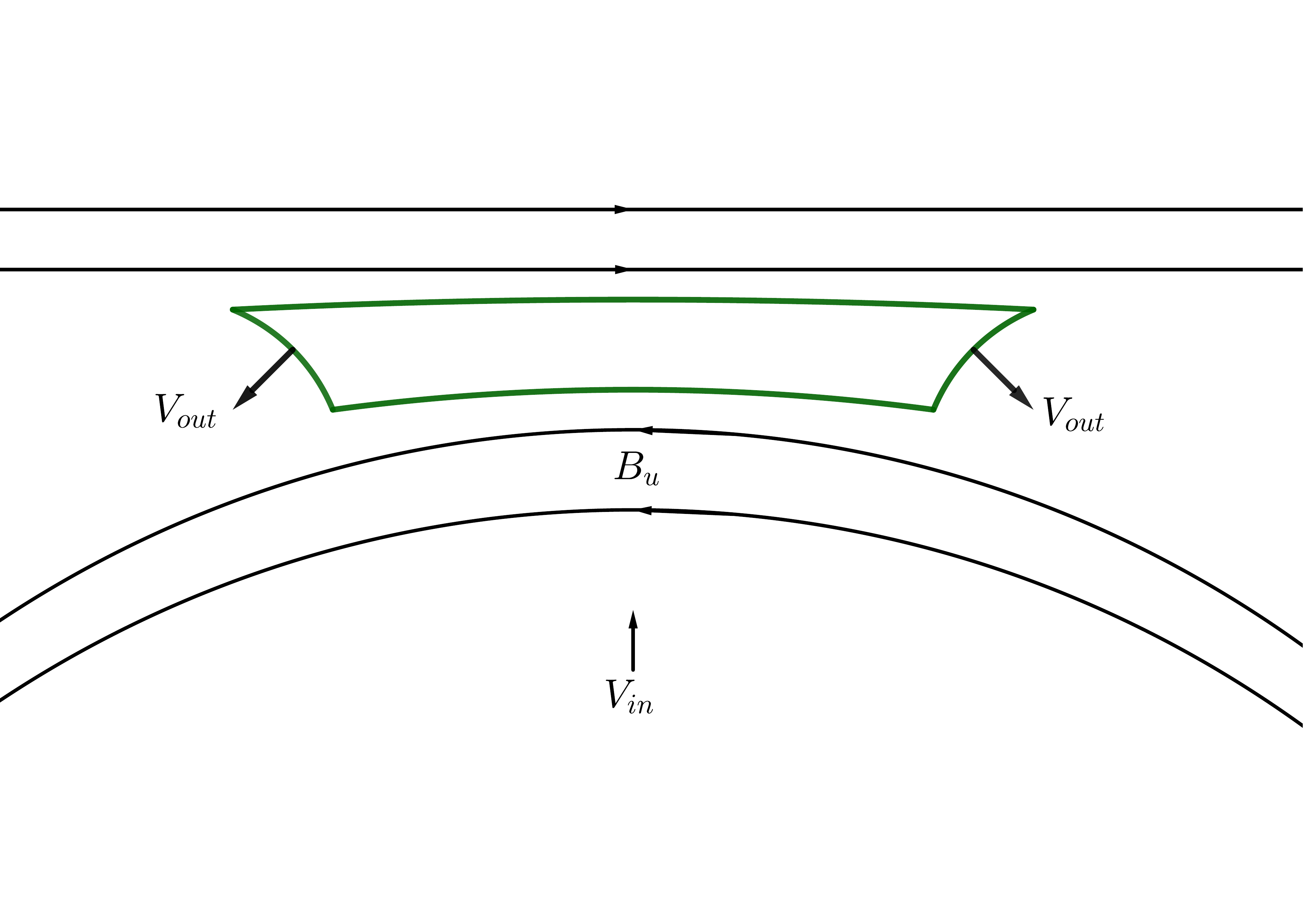}
  \caption{Schematic diagram showing the simplified geometry of magnetic field lines and of one of the two associated SP-like current sheets
  formed by the tilt instability.}
\label{fig8}
\end{figure}

\subsection{Formation of a few plasmoids close to the critical Lundquist number $S_c$}

Exploring now lower resistivity values, situated in a range $[5 \times 10^{-4} : 2.5 \times 10^{-4}]$, corresponding thus to $S^*$ in the range
$[2 \times 10^{3} : 4 \times 10^{3}]$, lead to estimated $S$ values in a range $[3100: 6700]$ close to $S_c$ from our simulations.
The time history of the maximum
current density is plotted in right panel of Figure 9 for $3$ different $S*$ values. First, we note that the formation of plasmoids occurs sooner when $S$
is higher. For example, for $S^* = 2857$, a first single plasmoid is seen to form close to the first peak ($t \simeq 8.5$ $t_A$). However, they can appear
earlier for $S^* = 4000$ or equivalently $S = 6700$ (i.e. before the first current peak) with two plasmoids invading each current sheet at $t \simeq 9.5$ $t_A$. For, the highest resistivity
case (i.e. for $S^* = 2000)$, a single plasmoid is seen to form very lately (i.e. at  $t \simeq 11.5$ $t_A$) on one current sheet, as also seen in the corresponding
left panel of Figure 9. The ability of our code to capture the plasmoid structure is illustrated in Figure 10, showing a zoom on current density with the associated adapted grid.
Note that, for the run employing $S^* = 2000$, the tilt instability has led to a sense of rotation of the initial setup in the opposite sense (i.e. counter clockwise) compared to the
run of Figure 2.

Thus we consider that the critical Lundquist number is $S_c \simeq 5 \times10^3$ for our tilt setup. This is a factor of two smaller than the currently admitted value of
$10^4$. However, this is also significantly lower than the critical Lundquist deduced from MHD simulations using the coalescence setup where
$S_c \simeq 3 \times10^4$ is reported \citep{hua10, hua17}.

\begin{figure}
\centering
 \includegraphics[scale=0.15]{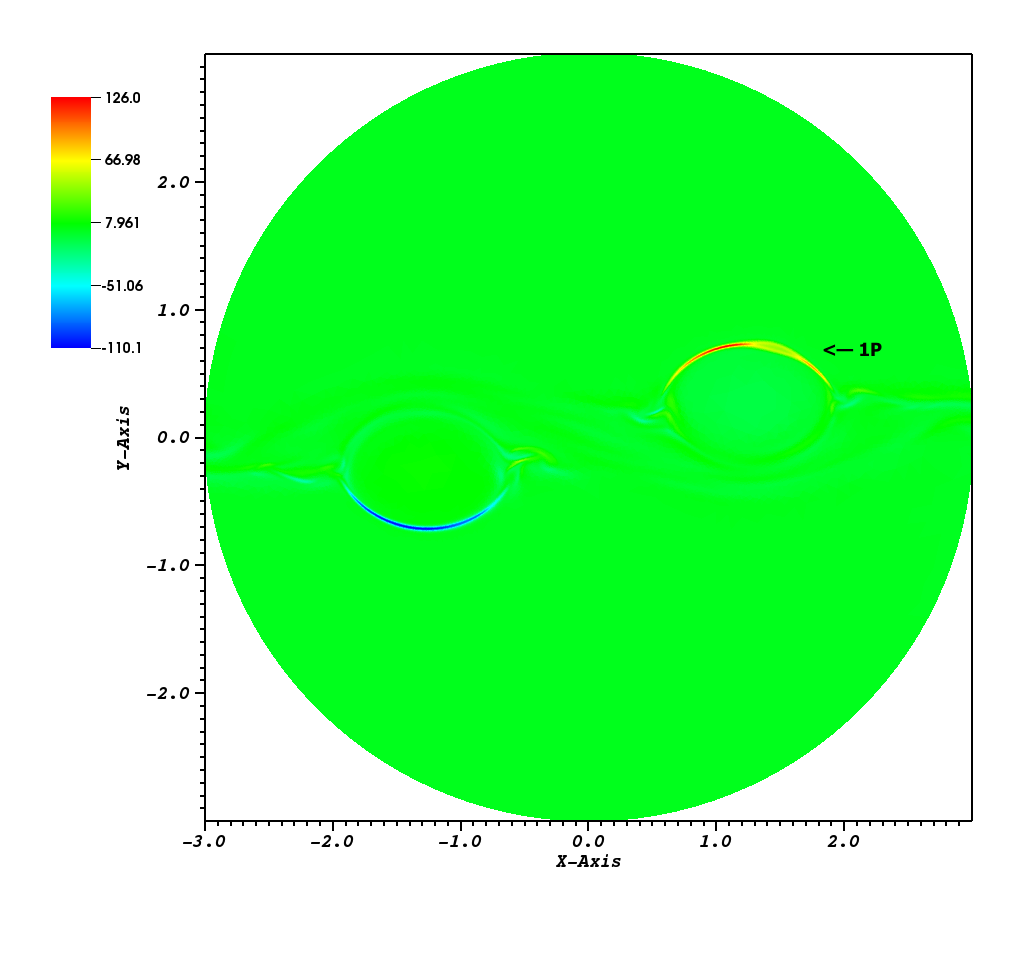}
 \includegraphics[scale=0.57]{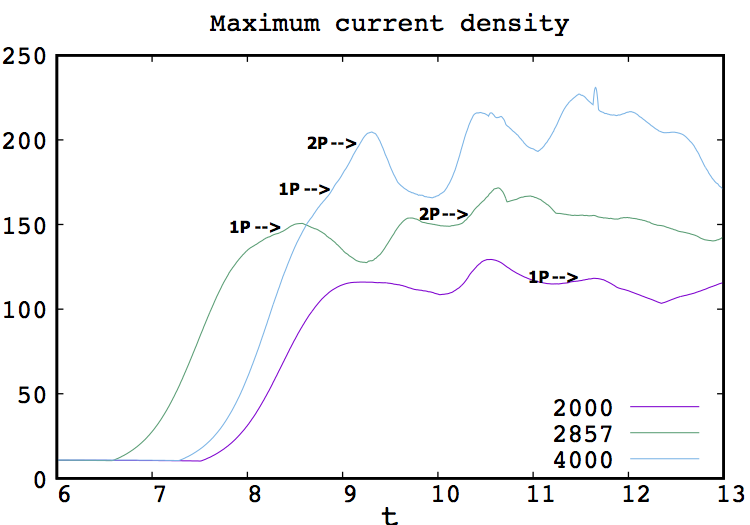}
  \caption{(Left panel) Colored contour map of the current density during reconnection phase at $t = 11.5$ $t_A$, for an run employing an
  inverse resistivity value $S^* = 2000$. A single plasmoid (label $1P$) is formed on the right current sheet.
(Right panel) Time history of the maximum current density for $3$ runs (employing $S^* = 2000,  2857, 4000$). The label $1P$ indicates
the early formation of a single plasmoid on one current sheet and $2P$ the presence of two plasmoids on the same sheet. }
\label{fig9}
\end{figure}

\begin{figure}
\centering
 \includegraphics[scale=0.18]{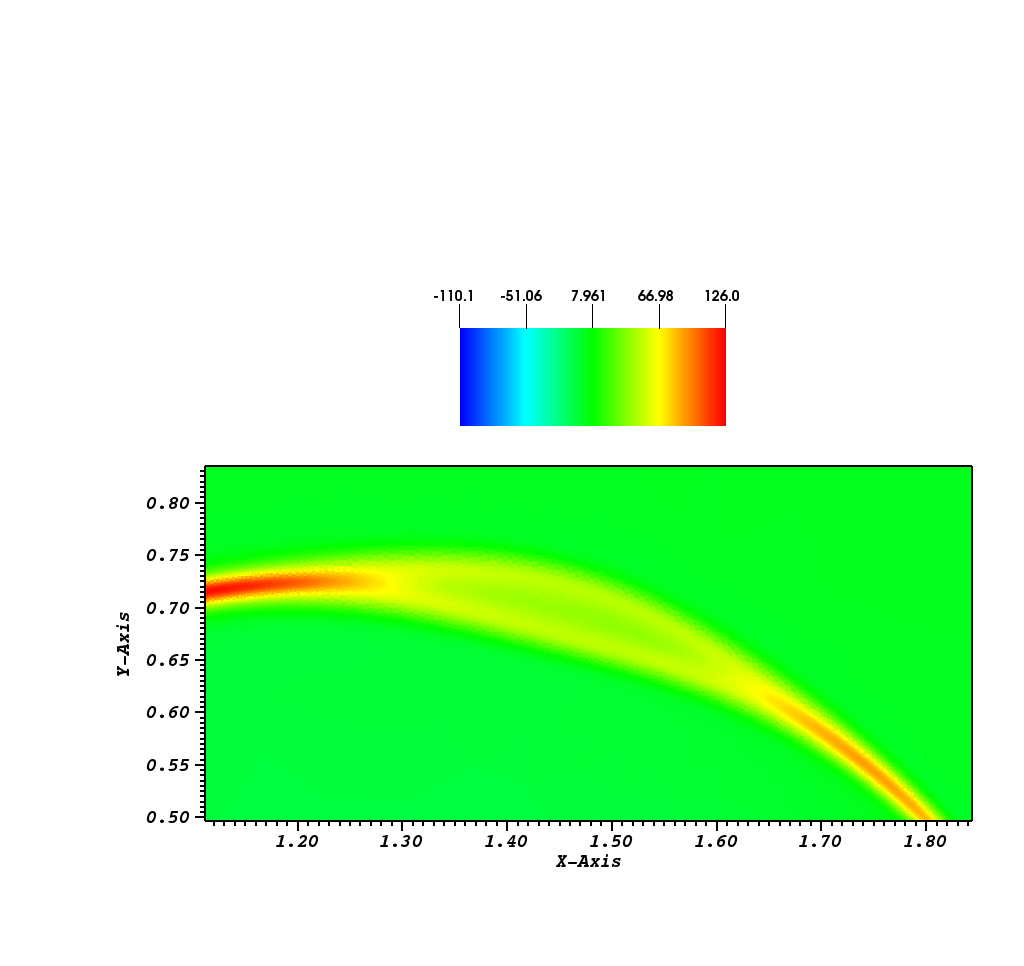}
 \includegraphics[scale=0.18]{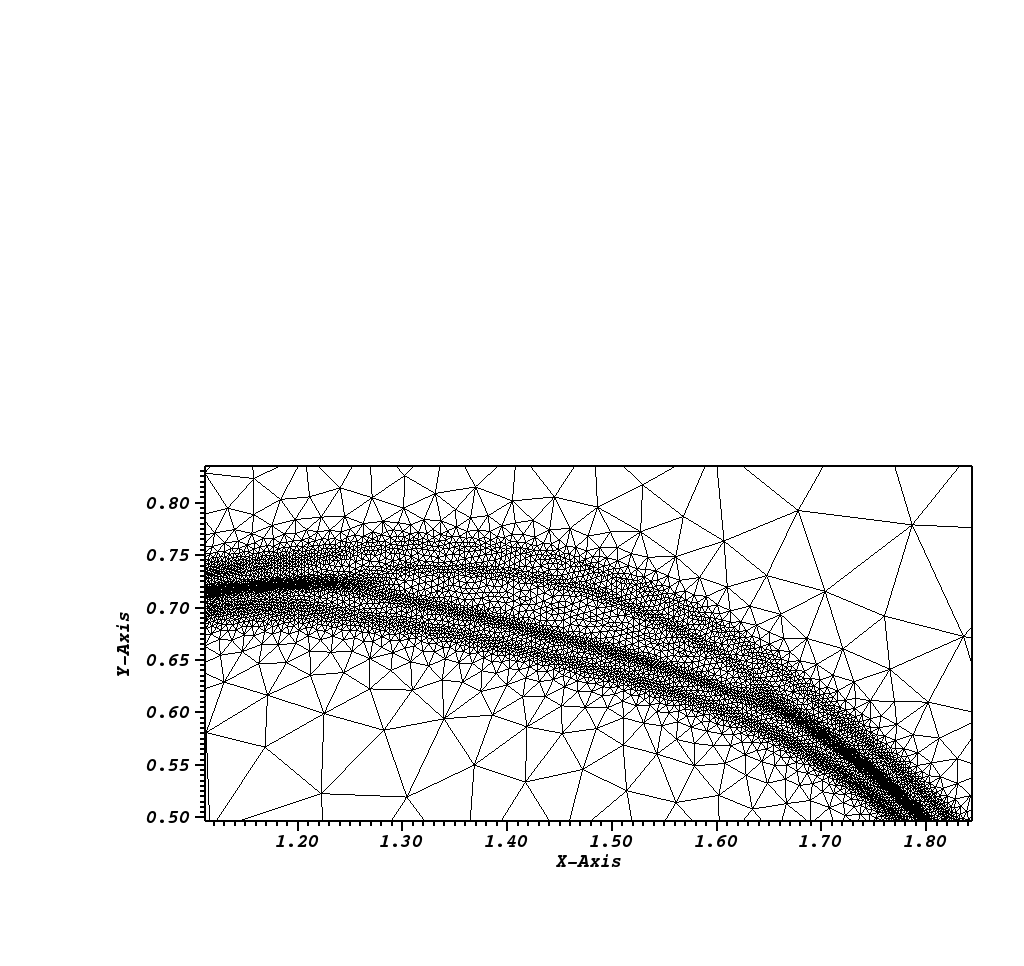}
  \caption{Zoom-in of the current density taken from the previous figure (left panel), and corresponding adapted mesh (right panel).}
\label{fig10}
\end{figure}

\subsection{Formation of many plasmoids in the high Lundquist regime and associated fast reconnection}

We focus on runs using higher Lundquist values, with $S^*$ in a range $[6  \times 10^3 : 5  \times 10^5]$, translating into
$S$ values in the approximated range $[1  \times 10^4 : 1  \times 10^6]$. The results of time evolution of the maximum current density
for a few runs in this $S$ range are reported in Figure 11. As illustrated on the curve (case for $S^* = 2.5  \times 10^4)$, $3$ asterisks
are used to indicate, the beginning of the tilt phase, the early formation of plasmoids, and the disruption of the current layers by
the plasmoids successively. At time spotted by the second asterisk, the plasmoids are barely visible as their amplitude remain smaller than
the (forming) current sheet contribution. Note that the second asterisk also coincides with an abrupt change
of slope in the current density. At time spotted by the third asterisk, the plasmoids are able to fully disrupt the current layers that
consequently lose their integrity, as shown in Figure 12 for two runs. Note that a contour map using a saturated value significantly
lower than the maximum current density is necessary to distinguish the plasmoids at disruption time.
The disruption is subsequently followed by a stochastic reconnection phase with time oscillations
of the maximum current density around an average value. We have observed that the number of plasmoids is maximum at
disruption and fluctuates during magnetic reconnection process, as new plasmoids are constantly forming, moving, eventually coalescing
(giving thus monster plasmoids), and finally being ejected through the two end layers.
The transition between a current layer slightly modified by the early growth of the plasmoids (upper panel) and a fully disrupted
layer (lower panel) is illustrated in Figure 13 for the run with $S = 3.5 \times 10^5$. The formation of secondary current sheets giving rise
to secondary plasmoids, and the coalescence effect between primary plasmoids leading to bigger plasmoids, are observed only
at a time following the disruption (i.e. at  time spotted by the third asterisk). This gives support to our assumption (see below) that the
phase between the second and third asterisk corresponds to a linear one for the plasmoids growth.
 At a given time during reconnection, the system thus appears as two aligned layer structures with a relatively high number of plasmoids ($10$ for
$S^* = 2.5  \times 10^4$, and $20$ for $S^* = 2 \times 10^5$ runs) of different sizes.

We can define two parameters characterizing the plasmoids development. The first one is the time needed for the first plasmoids
to appear once the current sheet begins to form, i.e. the time delay between the first two asterisks in Figure 11, which we call $t_p$.
The second one is deduced from the slope in current density observed between the second and third asterisk that can be fitted as
$e^{\gamma_p t}$. $\gamma_p$ can thus be interpreted as an instantaneous maximum growth rate. 
We have reported in Figure 14 (left panel), the values estimated from our
runs of the two previously defined parameters ($t_p$ and $\gamma_p$ ) as functions of $S$. 

\begin{figure}
\centering
 \includegraphics[scale=0.77]{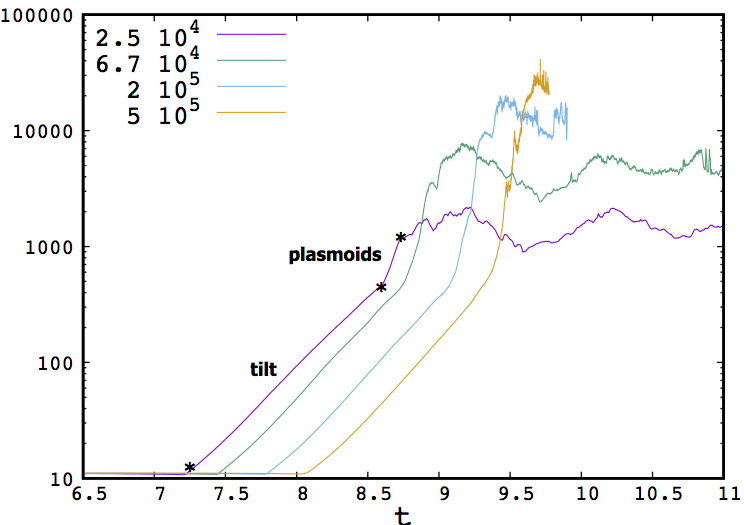}
  \caption{Time history of the maximum current density for different runs employing inverse resistivity values $S^*$ in the range
between $2.5  \times 10^4$ and $5  \times 10^5$. Note that for the two highest cases, only the early reconnection phase is
simulated. For the case employing the lowest $S^*$, Three asterisks are added indicating, the beginning of the tilt evolution, the early formation
of the first plasmoids, and the obtention of the stochastic reconnection regime (time at which a maximum number of plasmoids is observed)
successively. }
\label{fig11}
\end{figure}

\begin{figure}
\centering
 \includegraphics[scale=0.44]{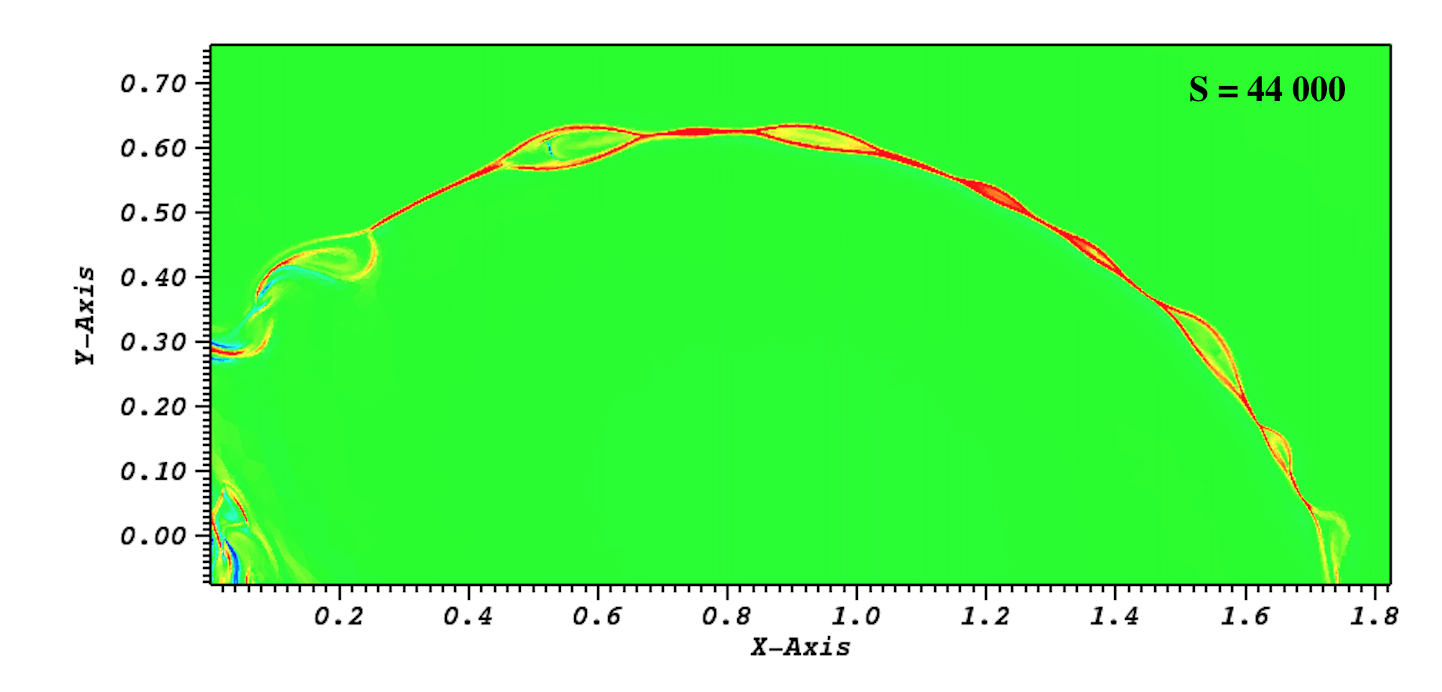}
 \includegraphics[scale=0.44]{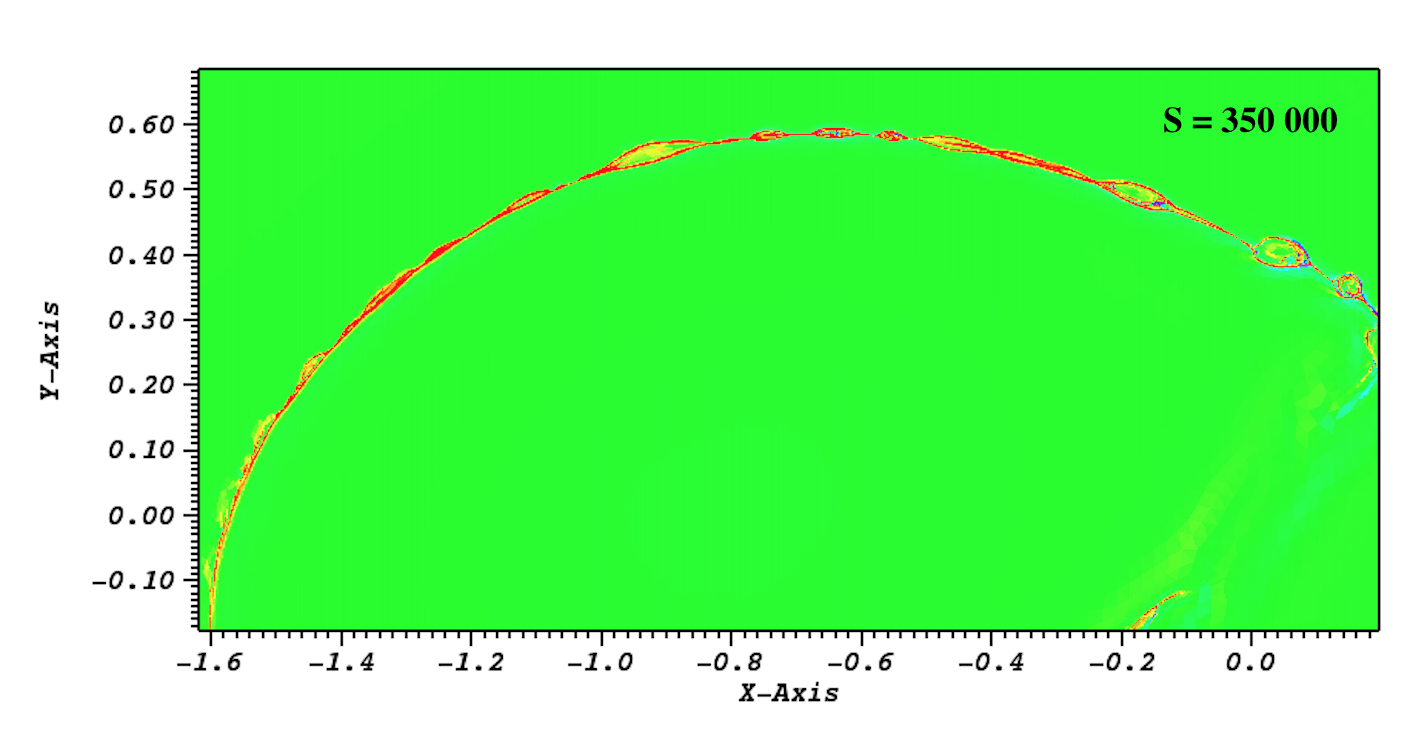}
  \caption{Zoom-in of one current sheet (colored contour map of current density with a scale saturated to $300$) obtained at an early time during the reconnection phase
 (i.e. at a time close to the third asterisk in previous figure) 
  for two runs employing inverse resistivity values $S^* = 2.5 \times 10^4$ (upper panel), and
 $S^* = 2 \times 10^5$ (lower panel). The corresponding evaluated $S$ numbers are indicated on the plots.}
\label{fig12}
\end{figure}

\begin{figure}
\centering
 \includegraphics[scale=0.44]{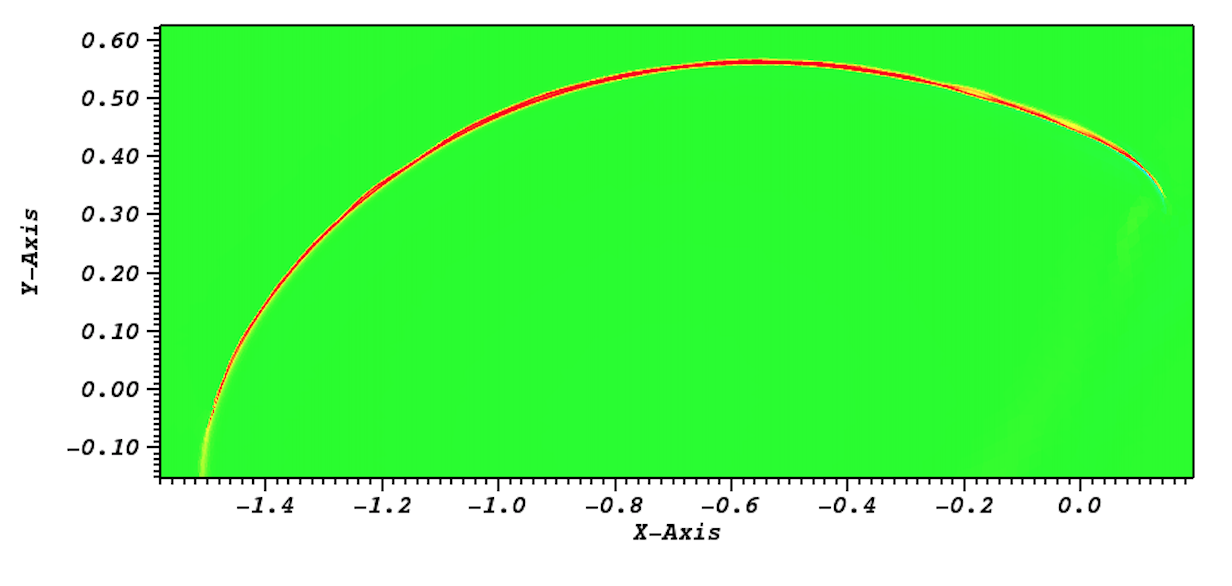}
 \includegraphics[scale=0.44]{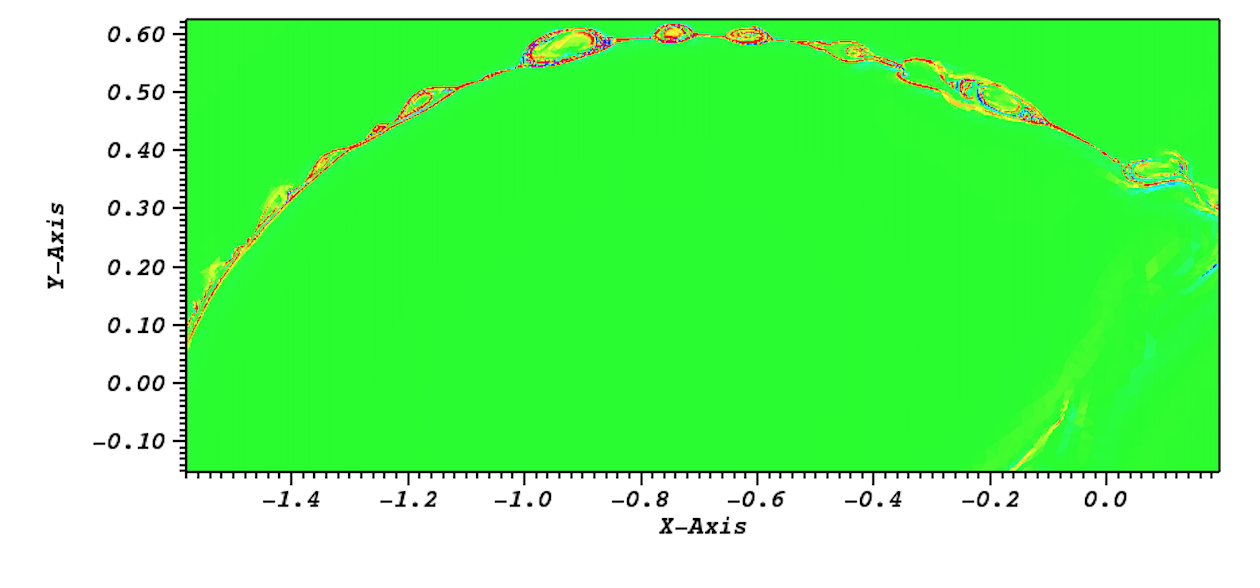}
  \caption{Same as previous figure for the case $S = 350000$, for the previous time spotted by the second asterisk in Figure 11 (upper panel), and for a slightly later time 
  during the reconnection process (lower panel). Note that the integrity of the current layer is not lost for the upper plot.} 
\label{fig13}
\end{figure}

First, one can observe that the characteristic time $t_p$ rapidly converges towards a value of $1.2$ $t_A \simeq 2.4$ $\tau_A$ that
becomes rapidly independent of $S$.
A more complex non monotonic dependence is obtained for the maximum growth rate. Indeed, $\gamma_p$ is shown to follow
a fitted scaling law $\gamma_p t_A = 0.9$ $S^{1/4} \simeq S^{1/4}$ only for a limited range of $S$ values between 
$6 \times 10^4$ and $2 \times 10^5$. Indeed, this agrees with the theoretical SP stability predicting $\gamma_{p} L/v_A = 0.62$ $S^{1/4} (1 + P_m)^{-5/8}$
\citep{com16}. For lower $S$ values, our $\gamma_p$ values are significantly lower than expected values from SP stability theory.
For the highest $S$ values, a transition towards a dependence scaling 
as $ K \ln ( C S^{-1/3})$ in agreement with the asymptotic solution of \citet{com16b} ($K$ and $C$ being constants, see Equation 19 in their paper). 
The value of $K$ is arbitrarily adjusted to unity, as it depends on different factors of order unity. This is not the case for the value of $C$ that is adjusted
to the very high value of $10^{10}$, as it is mainly determined by a small parameter $\omega_0$ representing the unknown noise.
 Thus, we can infer a noise level of order $10^{-10}$ in our simulations, as $C \sim 1/ \omega_0$.
The maximum value obtained for $\gamma_p$ is $\gamma_p t_A \simeq 20$, leading thus
to $\gamma_p \tau_A \simeq 10$. 

An additional parameter characterizing the plasmoids development is the maximum number of plasmoids $n_p$ (at a time
close to third asterisk in Figure 11). The results obtained for the different runs (right panel of Figure 14),
show again a convergence for intermediate $S$ values towards a scaling law, that is $n_p \simeq 0.2  \times S^{3/8}$, in close agreement with
the maximum wavenumber predicted from SP stability theory as $k_{p} L = 1.4 \times S^{3/8} (1 + P_m)^{-3/16}$,
using also $n_p = L k_p /(2 \pi)$. Again, for highest $S$ values, a transition to another dependence in agreement
with the asymptotic solution of \citet{com16b} is observed (see equation (18) where $k_p \propto S^{1/6} [\ln (C S^{-1/3})]^{5/6}$).
The highest value for the number of observed plasmoids is $n_p  \simeq 22$ for $S \simeq 10^6$.

Finally, we have plotted in Figure15 (left panel) the maximum current density obtained at saturation as a function of $S$ for all the
runs. The results clearly show a transition between two regimes at a critical Lundquist $S_c \simeq 5 \times 10^3$. 
Indeed, the values for lower $S$ values perfectly follow a Sweet-Parker scaling as $2 \times S^{1/2}$, while
another scaling increasing linearly with $S$ is obtained for higher $S$ values as $0.033 \times S$. A similar
results ensues for the maximum vorticity (right panel in Figure 15), where a transition between a SP scaling as $0.75 \times S^{1/2}$ and
a linear one scaling as $0.0125 \times S$ occurs.
As concerns the reconnection rate, constant values of $\eta J_{max}  \simeq 0.05$ (independent of $S$) is deduced
in the plasmoid regime corresponding to a normalized rate (by dividing by $V_A B_u$) of $0.014$.
This value is in very good agreement with the value expected form previous studies, and also from the theoretical
estimate of $ \epsilon_c  (1+ P_m)^{-1/2} \simeq 0.012$, where $\epsilon_c \simeq 1.7 \times 10^{-2}$ is deduced from
the expression $S_c = \epsilon_c^{-2} (1 + P_m)^{1/2}$ \citep{com16}.

\section{Comparison with theory and previous studies}

First, from our knowledge, this study is the first one to address into detail the reconnection process associated with
the nonlinear evolution of the tilt instability. Interesting results, even in the SP regime are obtained. Indeed, two forming twin
current sheets (with current density of opposite sign) drive a steady-state reconnection in agreement with classical scaling laws
given by the famous Sweet-Parker model.  A slight amendment (by a factor of two) for the vorticity of the outflow is however required
due to the particular asymmetry associated with the curved geometry of the current layers (see Figure 8).

In our study using the tilt instability as a triggering mechanism to form the current sheets,
the transition from a SP reconnection process to a plasmoid-dominated regime occurs for a critical Lundquist number
$S_c \simeq 5 \times 10^3$. This is a factor of two lower than the often-quoted $S_c \sim 10^4$ value in the literature. However,
there is no precise universal value, as it depends on different parameters like, the current sheet geometry (via the choice for the initial setup),
the magnetic Prandtl number, and also the noise amplitude (via the numerical scheme in our study).
The exact definition of the Lundquist number can also differ slightly from one study to another. 
In the case of the numerical study using the coalescence instability, a value of $S_c \sim 3  \times 10^4$ has
been reported in simulations assuming zero explicit viscosity.
Even for Lundquist number very slightly lower than $S_c$, the formation of a transient single plasmoid
is observed to occur at a relatively late time, with no real impact on the SP reconnection rate (see Figures 9-10).

Focusing on the plasmoid-dominated regime, we have examined the growth of the plasmoids, from their
birth to their ensuing disrupting effect on the current sheets. An important reference time scale for comparing the
latter growth is the time scale for forming the current layers, that is given by $\tau = 0.38$ $t_A \simeq 1$ $\tau_A$, as the tilt
mode is an ideal MHD instability leading to a current density increasing exponentially as $e^{2.6 t}$.
This triggering phase in our simulations has been carefully checked to agree with stability theory
\citep{ri90}. As seen in Figure 6, the formation of the current sheets proceeds trough a combination
of thinning (as $a$ is observed to decrease in time), stretching (as $L$ is increasing), and a weak magnetic field 
strengthening, in agreement with the scenario suggested by Tolman et al. (2019).

We have defined two simple parameters characterizing the growth of the forming plasmoids.
The first one, $t_p$, is the delay time between the birth of the first plasmoids (time at which they become
to be barely visible in the current density structure) and the start of the formation of the currents sheets (taken as the time at which the
corresponding current density exceeds the equilibrium setup value).
A rapidly converged constant value (with $S$) of $t_p = 2.4$ $\tau_A$ is obtained (see Figure 14).
This delay time can be identified to correspond to the quiescent phase proposed in Comisso et al.'s scenario, during which many modes become 
progressively unstable and compete with each other (see Figures 3-4 in \citet{com17}).
Indeed, the duration of this phase is predicted to be approximatively given
by the time scale of the current sheet formation. This is also in agreement with a conclusion drawn in \citet{uzd16}.
A similar result has been obtained for the coalescence setup, with a slight difference for the highest $S$ values
where their time delay is non-monotonic and increases weakly again \citep{hua17}. 

The second parameter, $\gamma_p$, is deduced from the second slope observed during the
increase of the maximum current density (see Figure 11), and thus characterizes an abrupt growth phase 
following the slower previous quiescent phase. This explosive phase
over a short time scale corresponds to the predicted phase dominated
by the mode that emerges "first" at the end of the linear phase in the theory of Comisso et al.
Our results also qualitatively agree with the non-monotonic dependence with $S$, as a consequence of the non-power
law dependence with $S$. Values $\gamma_p  \tau_A \simeq 10$ are also obtained 
for the highest $S$ values, thus confirming that $\gamma_p  \tau_A >> 1$ at the end of the linear phase.
For $S  \simgt S_c$, the scaling law given by SP stability theory with $\gamma_p  \propto S^{1/4}$ is only marginally recovered.
A very similar result has been obtained for the coalescence setup \citep{hua17}, 
except that $\gamma_p  \tau_A \simeq 10-20$ are reported. 
The difference at relatively low $S$ can be largely attributed to the reconnection outflow (neglected in theoretical models) that can affect the growth of the plasmoids
and thus the scaling relations (Huang et al. 2019).
As shown in previous studies, the noise induced by the numerical simulations also influences the results, and thus
is an important parameter that needs to be investigated in future studies.

However, our results seem to contradict predictions from model of \citet{puc14}, where the linear growth of
plasmoids is constant and at most Alfv\'enic. This latter model is based on the existence of the critical aspect ratio
$L/a \simeq S^{-1/3}$ for the disruption of the current sheet. We have measured the final aspect ratio just prior to
the disruption in our runs, as plotted in Figure 16. Our results clearly show that the SP aspect ratio is reached
only at low $S$, and final aspect ratio $L/a$ turns to be bounded as $S^{1/3} < L/a < S^{1/2}$ at higher $S$. This is
in agreement with the predicted bounding domain $ \tau^{-2/3} S^{1/3} < L/a < S^{1/2}$ \citep{com17}.

Finally, our estimate of the time-averaged normalized reconnection rate is $0.014$, that is two times higher than the value deduced
from the coalescence setup. It is nevertheless in good agreement with values obtained in the literature of $\sim 0.01$ much higher
than the Sweet-Parker rates, which could be sufficient to explain many disruptive events if the collisionnel regime apply.
A fractal model (with hierarchical structure of the plasmoid chains that are effectively observed in simulations) based on heuristic arguments
has been proposed to explain this fast rate independent of the Lundquist number \citep{hua10}. Indeed, to this end, a number of plasmoids
(called non linear number) is required to scale linearly with $S$ \citep{hua10}. Investigating this point is a complicated task
requiring longer time simulations, and it will be the subject of a future study using tilt setup.

\begin{figure}
\centering
 \includegraphics[scale=0.50]{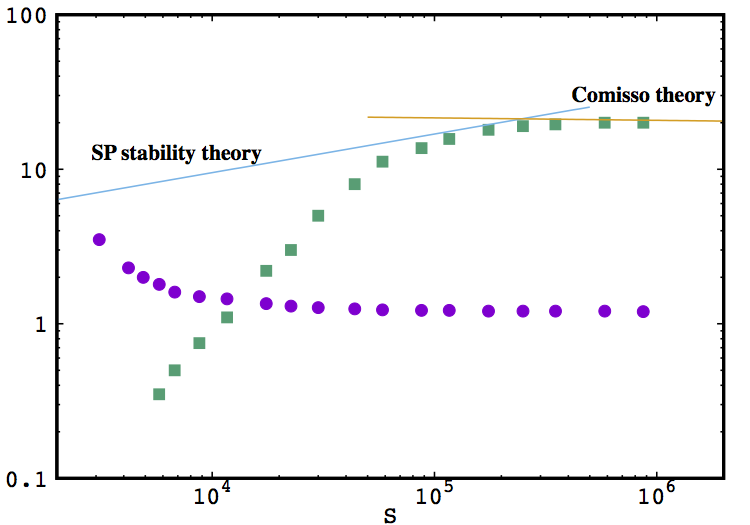}
 \includegraphics[scale=0.50]{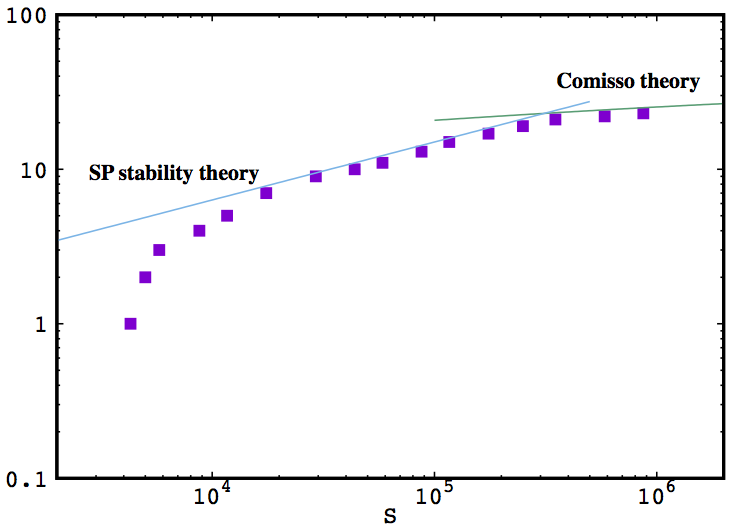}
  \caption {The two parameters characterizing the early growth of the plasmoids, $t_p$ (in units of $t_A$, circles) and $ \gamma_p$
  (in units of $1/t_A$, squares) as a function of $S$ (left panel). The corresponding maximum number of
  plasmoids $n_p$ (right panel), obtained at a time spotted by the third asterisk in previous figure. Theoretical power laws
  expected from SP linear theory scaling as $0.9 \times S^{1/4}$ and $0.2 \times S^{3/8}$, and asymptotic solutions from Equations (18-19)
  of \citep{com16b} are also plotted for comparison (see also text).
  } 
\label{fig14}
\end{figure}

\begin{figure}
\centering
 \includegraphics[scale=0.50]{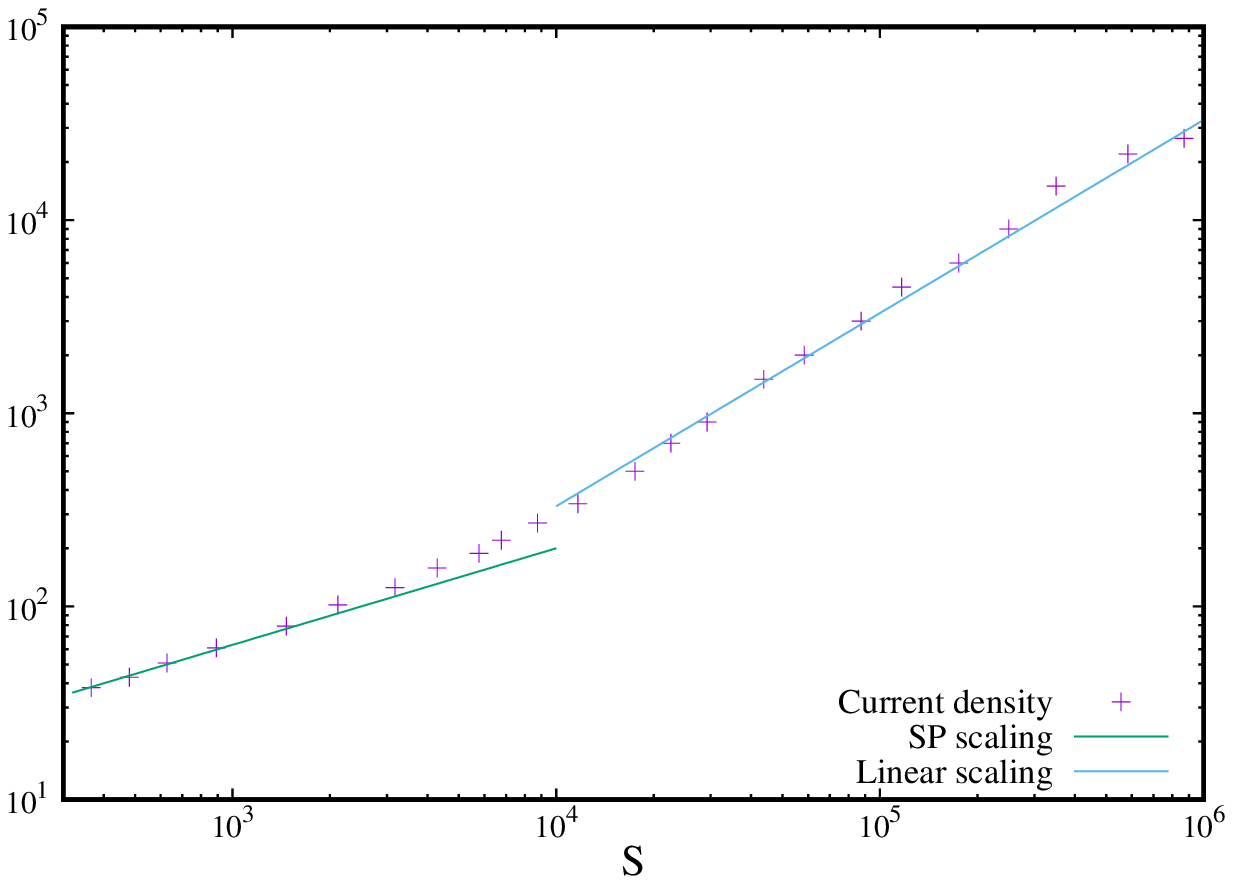}
 \includegraphics[scale=0.50]{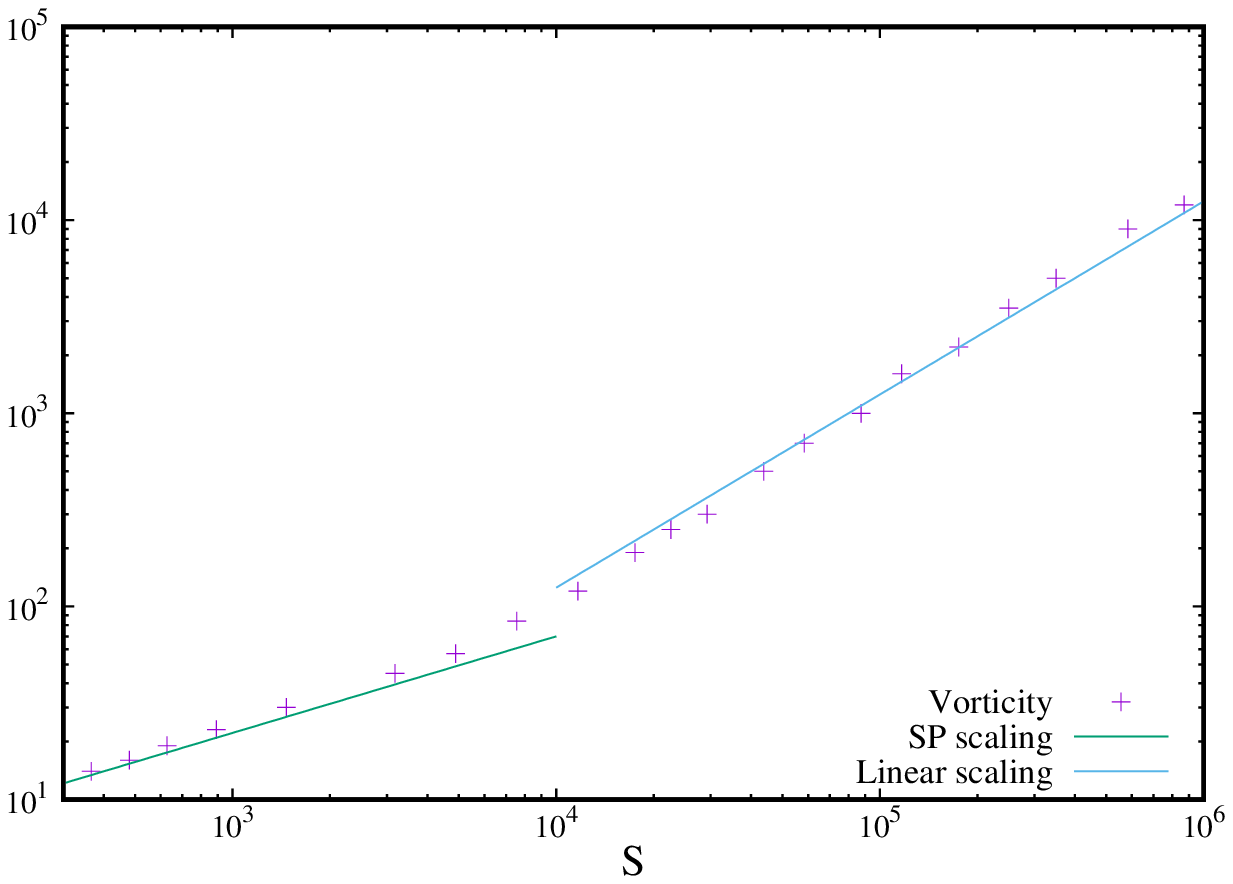}
  \caption{Overview of the maximum current density (left panel) and maximum vorticity (right panel) as a function of the Lundquist
 number $S$, for all the different runs. SP scaling laws in $2 \times S^{1/2}$ (for current density) and  $0.75 \times S^{1/2}$ (for vorticity)
 are plotted, as well as linear scaling laws in $0.033 \times S$ (for current density) and $0.0125 \times S$ (for vorticity) approximating
 the plasmoid-dominated regime.}
\label{fig15}
\end{figure}

\begin{figure}
\centering
 \includegraphics[scale=0.7]{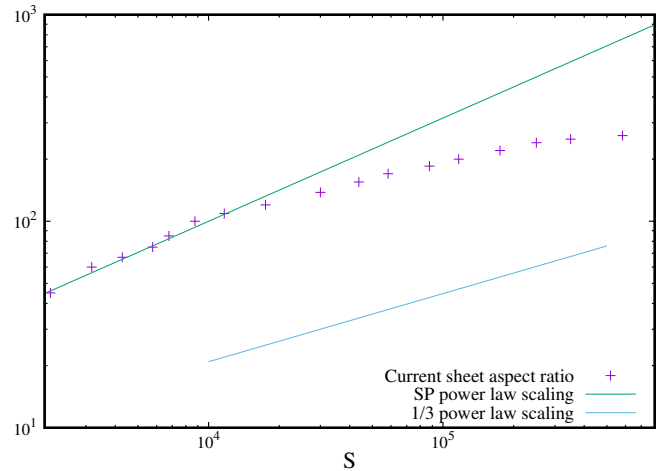}
  \caption{Current sheet aspect ratio values $L/a$, in SP reconnection regime for $S \simlt  5 \times 10^3$, and just before disruption in plasmoid-dominated regime for higher $S$ values.
  Theoretical power laws scaling as $S^{1/2}$ and $S^{1/3}$ are plotted for comparison.}
\label{fig16}
\end{figure}

\section{Conclusion}

In this study we have demonstrated the usefulness to consider other configurations than the single Harris-type
configuration or the coalescence instability to study the onset of the plasmoid-dominated reconnection regime
in forming current sheets. Our results being very similar to the those obained from the coalescence setup, suggest that Comisso et al.'s model
is able to correctly predict the explosive growth of plasmoids leading to disruption of the reconnection current sheets when the
initial configuration is ideally unstable. On the other hand, the other model developped by \citet{puc14} could apply when the
initial configuration is ideally stable (and thus resistively unstable), as it has been validated using Harris-type current layer.
This could explain the fastest time scale involved in the first category of setup compared to the second one. 

The Lundquist number reached in this study is high enough in order match the relevant values for
tokamaks. Indeed, the relevant $S$ value for the internal disruption associated with the internal kink mode is
$S  \simeq 10^5$, as $S = 0.004 S^*$ ($S^* = 2.5$ $ 10^7$ being a standard Lundquist number value defined
in terms of the toroidal magnetic field)  \citep{gun15}. The corresponding width of the Sweet-Parker current layer is thus estimated to be $a \simeq 1$ cm,
and the smallest length scale associated to the plasmoid structure is probably of order $1$ mm or even smaller, reaching
thus a scale close to the the kinetic ones. Kinetic effects could be incorporated to our model in order to address this
point. For example, the plasmoid instability has been shown to facilitate the transition to a Hall reconnection 
in Hall magnetohydrodynamical framework with an even faster reconnection rate of $\sim 0.1$ \citep{hua11}.

The smallest length scale associated to the plasmoid structure for $S = 10^6$ remains larger than the kinetic scale 
that is of order $10$ m, when considering a solar loop structure and taking a length $L = 10^7$ m. However, as very high
Lundquist number (at least $10^{10}$) is required for the solar corona, kinetic effects could also play a role if the kinetic scale
is reached via the plasmoid cascade at such huge Lundquist number.


\bibliographystyle{jpp-tilt}

\end{document}